\documentclass[12pt]{article}
\usepackage{graphics,scipp,equations}

\begin{document}
\setlength{\unitlength}{1mm}  

\pagestyle{empty}
\topmargin-2pc
\begin{flushright}
SCIPP 98/47     \\
August, 1999 \\
hep--ph/9909335
\end{flushright}
\vskip2cm

\begin{center}
{\Large\bf  Radiative Corrections to the \boldmath$Z b \bar{b}$ Vertex \\[6pt]
 and Constraints on Extended Higgs Sectors}\\[1cm]
{\large Howard E. Haber and Heather E. Logan}\\[3pt]
{\it Santa Cruz Institute for Particle Physics  \\
   University of California, Santa Cruz, CA 95064, U.S.A.} 
\\[1.5cm]

{\bf Abstract}
\end{center}

We explore the radiative corrections to the process $Z \rightarrow b
\bar{b}$ in models with extended Higgs sectors.  The observables
$R_b\equiv \Gamma(Z \rightarrow b \bar{b})/\Gamma(Z \rightarrow
\mathrm{hadrons})$ and the $Zb\bar b$ coupling asymmetry, $A_b\equiv
(g\ls L^2 - g_R^2)/(g\ls L^2 + g_R^2)$, are sensitive to these
corrections.  We present general formulae for the one-loop corrections
to $R_b$ and $A_b$ in an arbitrary extended Higgs sector, and derive
explicit results for a number of specific models.  We find that in
models containing only doublets, singlets, or larger multiplets
constrained by a custodial $SU(2)_{c}$ symmetry so that $M_W = M_Z
\cos\theta_W$ at tree level, the one-loop corrections due to virtual
charged Higgs bosons always worsen agreement with experiment.  The
$R_{b}$ measurement can be used to set lower bounds on the charged
Higgs masses.  Constraints on models due to the one-loop contributions
of neutral Higgs bosons are also examined.

\vskip2cm


\vfill
\clearpage

\makeatletter
\def\@cite#1#2{{[{#1}]\if@tempswa\typeout
{IJCGA warning: optional citation argument
ignored: `#2'} \fi}}


\newcount\@tempcntc
\def\@citex[#1]#2{\if@filesw\immediate\write\@auxout{\string\citation{#2}}\fi
  \@tempcnta\z@\@tempcntb\m@ne\def\@citea{}\@cite{\@for\@citeb:=#2\do
    {\@ifundefined
       {b@\@citeb}{\@citeo\@tempcntb\m@ne\@citea\def\@citea{,}{\bf ?}\@warning
       {Citation `\@citeb' on page \thepage \space undefined}}%
    {\setbox\z@\hbox{\global\@tempcntc0\csname b@\@citeb\endcsname\relax}%
     \ifnum\@tempcntc=\z@ \@citeo\@tempcntb\m@ne
       \@citea\def\@citea{,}\hbox{\csname b@\@citeb\endcsname}%
     \else
      \advance\@tempcntb\@ne
      \ifnum\@tempcntb=\@tempcntc
      \else\advance\@tempcntb\m@ne\@citeo
      \@tempcnta\@tempcntc\@tempcntb\@tempcntc\fi\fi}}\@citeo}{#1}}
\def\@citeo{\ifnum\@tempcnta>\@tempcntb\else\@citea\def\@citea{,}%
  \ifnum\@tempcnta=\@tempcntb\the\@tempcnta\else
   {\advance\@tempcnta\@ne\ifnum\@tempcnta=\@tempcntb \else \def\@citea{--}\fi
    \advance\@tempcnta\m@ne\the\@tempcnta\@citea\the\@tempcntb}\fi\fi}
\@addtoreset{equation}{section}
 \def\theequation{\thesection.\arabic{equation}}
\makeatother
\def\ifmath#1{\relax\ifmmode #1\else $#1$\fi}
\def\quarter{\nicefrac{1}{4}}
\def\half{\nicefrac{1}{2}}
\def\lsim{\mathrel{\raise.3ex\hbox{$<$\kern-.75em\lower1ex\hbox{$\sim$}}}}
\def\gsim{\mathrel{\raise.3ex\hbox{$>$\kern-.75em\lower1ex\hbox{$\sim$}}}}


\def\nicefrac#1#2{\hbox{${#1\over #2}$}}


\pagestyle{plain}

\section{Introduction}

The Standard Model of electroweak interactions has been 
tested to unprecedented precision during the past decade
at the LEP and SLC colliders \cite{Clare99,Mnich99,Morris99}.
Global fits of electroweak observables
have confirmed that the electroweak interactions are well 
described by a spontaneously broken SU(2) $\times$ U(1) gauge theory.
However, these measurements have not yet revealed the underlying
dynamics responsible for electroweak symmetry breaking (EWSB).

In the Standard Model (SM), the electroweak symmetry is broken by the
dynamics of a weakly coupled scalar Higgs sector consisting of
one complex SU(2) doublet of scalar fields with hypercharge $Y=1$.  
After EWSB, three scalar degrees of freedom (Goldstone bosons) are
absorbed by the $W$ and $Z$,
leaving one CP--even neutral Higgs boson $H^0$ in the physical
spectrum.   The SM Higgs sector possesses an unbroken global 
SU(2) symmetry of the EWSB sector, often called ``custodial SU(2)
symmetry'' \cite{Sikivie80}.  This symmetry leads to the tree-level
relation, $\rho\equiv M_W^2/M_Z^2 \cos^2 \theta_W=1$, a relation that is 
satisfied experimentally to better than a fews parts in a 
thousand \cite{Erler99}. 

Precision electroweak data is now accurate enough to provide
non-trivial tests of the one-loop structure of the SM.
In particular, one can begin to test the EWSB sector of the theory by
probing the one-loop virtual effects of the Higgs sector.     
The couplings of Higgs bosons to fermions and gauge bosons are 
proportional to the fermion and gauge boson masses, respectively.
As a result, one--loop corrections 
involving Higgs bosons coupled to $W$, $Z$ or third--generation quarks 
can be significant.  In the SM, loop corrections involving $H^0$ coupling
to gauge bosons depend logarithmically on the $H^0$ mass.  A 
fit to the electroweak data gives an upper bound on the SM Higgs mass of
$M_{h^0}\lsim 220$ GeV at the 95\% confidence level \cite{Clare99,Morris99}.
In the SM, the Higgs couplings to third--generation quarks do not provide
additional constraints on the Higgs sector.  Virtual Higgs exchange does
contribute to the decay $Z \rightarrow b \bar{b}$; however,
the coupling of $H^0$ to $b$-quarks is too small to make an 
observable contribution.  The coupling of the
charged Goldstone bosons $G^{\pm}$ to $t \bar{b}$ is large enough to 
make an observable contribution to $Z \rightarrow b \bar{b}$, but 
this contribution
is fixed by electroweak symmetry; it depends only on the $W$ and $t$--quark
masses, the electromagnetic coupling and $\sin^2 \theta_W$
\cite{Akhundov86,Beenakker88,Bernabeu88,Hollik90,Lynn90}. 

Many extensions to the minimal SM Higgs sector are possible.  (For a
comprehensive review, see ref.~\cite{HHG}.)  As in the SM, extended
models typically must contain at least one complex $Y=1$ SU(2) doublet
in order to give mass to the fermions.  Additional SU(2) doublets,
singlets, and/or larger multiplets may also be present.  Such
extended Higgs sectors contain charged Higgs bosons and/or additional
neutral Higgs bosons in the physical spectrum.  Some constraints on
the model exist due to the observed $\rho\simeq 1$; this can restrict
the choices of Higgs multiplets or require a fine-tuning of
the vacuum expectation values of the neutral Higgs fields.  In
addition, the experimentally observed suppression of flavor changing
neutral currents (FCNC's) implies that Higgs-mediated
tree-level FCNC's are either absent
(which constrains the Higgs-fermion couplings of the
model \cite{Glashow77,Paschos77}), or suppressed \cite{Atwood97}.
In the latter case, the suppression of FCNC's can be achieved
if the non-minimal Higgs states are sufficiently heavy 
(thereby approximately decoupling from the
sector of SM particles \cite{nirhaber}).

Extended Higgs sectors also contribute virtually to one-loop processes
involving SM particles.  In this paper our primary focus concerns the
electroweak observables associated with $Z\to b\bar b$.  In this case,
the Higgs sector can yield observable corrections at one-loop
through charged Higgs couplings to $ t \bar{b} $ and the 
neutral Higgs couplings to $ b \bar{b} $.  These can then provide new
constraints on the possible structure of the non-minimal Higgs sector.

The process $Z \rightarrow b \bar{b}$ yields two 
observable quantities, $R_b$ and $A_b$.
$R_{b}$ is the hadronic 
branching ratio of $Z$ to $b$ quarks,
\begin{equation}
                        R_{b} \equiv \frac{\Gamma(Z \rightarrow b \bar{b})}
                        {\Gamma(Z \rightarrow {\rm hadrons})}\,,
\end{equation}
and $A_b$ is the $b$-quark asymmetry,
\begin{equation}
A_b = \frac{\sigma(e^-_L \to b_F) - \sigma(e^-_L \to b_B)
                        + \sigma(e^-_R \to b_B) - \sigma(e^-_R \to b_F)}
                        {\sigma(e^-_L \to b_F) + \sigma(e^-_L \to b_B)
                        + \sigma(e^-_R \to b_B) + \sigma(e^-_R \to b_F)}\,,
\end{equation}
where $e^-_{L,R}$ are left and right handed initial--state electrons
and $b_{F,B}$ are final--state $b$-quarks moving in the forward and 
backward directions with respect to the direction
of the initial--state electrons.
In terms of the $b$-quark couplings to $Z$,
\begin{equation}
           A_{b} = \frac{(g^{L}_{Zb\bar{b}})^{2} - (g^{R}_{Zb\bar{b}})^{2}}
              {(g^{L}_{Zb\bar{b}})^{2} + (g^{R}_{Zb\bar{b}})^{2}}\,.
\end{equation}

In this paper we introduce a parameterization for a general extended Higgs 
sector and calculate the contribution to $Z \rightarrow b \bar{b}$ from 
one-loop radiative corrections involving singly charged and neutral Higgs
bosons.  
We obtain general expressions for the corrections to the left- and 
right-handed $ Z b \bar{b} $ couplings, and then use the measurements of 
$R_b$ and $A_b$ to constrain specific models.
This approach has the advantage of yielding general formulae for the 
corrections in terms of the couplings and masses of the Higgs bosons.
The formulae can then be specialized to any extended Higgs model by 
inserting the appropriate couplings.  Kundu and Mukhopadhyaya 
\cite{Kundu96} have taken the same approach and calculated the 
charged Higgs boson contributions to $Z \rightarrow b \bar{b}$ in a general
extended Higgs sector.  However, the neutral Higgs boson contributions in a 
general extended Higgs sector do not appear in the literature.
Specific results for the
one-loop corrections to $Z \rightarrow b \bar{b}$ in 
two-Higgs-doublet models (2HDM's) can be found in
refs.~\cite{Denner91,Djouadi91,Boulware91,Grant}.


One-loop corrections to $Z\to b\bar b$ can also arise from other
sources of new physics.  Thus, any derivation of constraints on the
Higgs sector based on the effects of Higgs virtual corrections must
assume that these are the dominant (or only) source of corrections
beyond the Standard Model.  For example, in theories of low-energy
supersymmetry, it is easy to find ranges of parameter space in which
the effects of virtual supersymmetric particle exchange compete (and
sometimes cancel out \cite{cancelout}) the effects of virtual Higgs exchange.
However, in the limit of large superpartner masses, the supersymmetric
contributions decouple \cite{Haber93,Herrero98},
and the formulae obtained in this paper are once again
applicable.

This paper is organized as follows.
In section~\ref{sec:data_constr} we discuss the measurements of 
$R_b$ and $A_b$ 
and the constraints that they put on the $ Z b \bar{b} $ couplings.
In section~\ref{sec:extendedHiggs} we
introduce the two Higgs doublet model and then generalize to an 
arbitrary extended Higgs sector.  
We then compute the radiative corrections to the $Zb\bar{b}$
coupling due to the virtual exchange of charged Higgs bosons
(section~\ref{sec-H+}) and neutral Higgs bosons 
(section~\ref{sec-H0}), respectively.
In section~\ref{sec:models} we apply the general formulae
for loop corrections to a number of specific models.  Based on the
current experimental measurements of $R_b$ and $A_b$, we exhibit the
constraints on the parameters of the extended Higgs sector.  
We first consider extended
Higgs sectors containing only doublets and singlets, and then extend
the analysis to Higgs sectors which containing larger multiplets in
addition to doublets.  Finally, we summarize our conclusions in
section~\ref{sec:conclusions}.  Additional details can
be found in ref.~\cite{hlthesis}.

\section{Constraints from the data}
\label{sec:data_constr}

The radiative corrections to $Z\rightarrow b \bar{b}$
modify the $ Z b \bar{b} $ couplings from their tree-level values.
In this section we show how the experimental constraints on $R_b$ and 
$A_b$ constrain the possible values of the effective $Zb\bar{b}$ couplings.
We employ the following notation for the effective $Zb\bar b$
interaction:
\begin{eqnarray}
{\cal L}_{Zb\bar b}&=&{-e\over 2s_W c_W} Z_\mu \bar b\gamma^\mu
\left[{\bar g}_b^L(1-\gamma_5)+{\bar
g}_b^R(1-\gamma_5)\right]b\nonumber \\
&=&{-e\over 2s_W c_W} Z_\mu \bar b\gamma^\mu({\bar v}_b-{\bar a}_b\gamma_5)b
\,,
\end{eqnarray}
where $s_W\equiv\sin\theta_W$ and $c_W\equiv\cos\theta_W$.
The effective couplings are then written as
\begin{equation}
    \bar{g}_b^{L,R} = g_{Zb\bar{b}}^{L,R} + \delta g^{L,R}\,,
\end{equation}
where $\bar{g}_b^{L,R}$ are the radiatively--corrected effective 
couplings, and the tree-level couplings are given by $g_{Zb\bar{b}}^L \equiv 
-\frac{1}{2} + \frac{1}{3}s^2_W$ and
$g_{Zb\bar{b}}^R \equiv \frac{1}{3}s^2_W$.



\subsection{Extracting the effective \boldmath$Zb\bar{b}$ couplings from \boldmath$R_b$
and \boldmath$A_b$}
\label{sec:gfromRbAb}

Following the discussion by Field \cite{Field98} and using his notation,
the effective couplings $\bar{g}_b^{L,R}$ are related to $R_b$ and $A_b$
as follows.
\begin{eqnarray}
R_b &=& \left[ 1 + \frac{S_b}{\bar{s}_b C_b^{QCD} C_b^{QED}} \right]^{-1}\,, 
                        \nonumber \\[6pt]
A_b &=& \frac{2 \bar{r}_b (1-4\mu_b)^{1/2}}
                        {1 - 4\mu_b + (1+2\mu_b)\bar{r}_b^2} \,,
\end{eqnarray}
where $C_b^{QCD}$ and $C_b^{QED}$ 
are QCD and QED radiative correction factors.
Using  $\alpha_s(M_Z) = 0.12$ and $\alpha^{-1}(M_Z) = 128.9$, the
numerical values of these factors are: $C_b^{QCD}=0.9953$
and $C_b^{QED}=0.99975$.  In addition,
\begin{eqnarray}
\bar{r}_b &\equiv& \frac{\bar{v}_b}{\bar{a}_b}\,,  \nonumber \\
\bar{s}_b &\equiv& (\bar{a}_b)^2(1-6\mu_b) + (\bar{v}_b)^2\,, \nonumber \\
S_b &\equiv& \sum_{q \neq b,t}\left[(\bar{a}_q)^2 + (\bar{v}_q)^2\right]\,,
              \nonumber \\
\mu_b &\equiv& \left[m_b(M_Z)/M_Z\right]^2\,.
\end{eqnarray}

In the definition of $S_b$, the sum is taken only over first and
second generation quarks.  To a good approximation, we can neglect
the contributions of new physics
to $S_b$, and fix this quantity to its SM predicted value.
Using the corresponding SM predicted values: ${\bar v}_u=0.1916$, 
${\bar a}_u=0.5012$, ${\bar v}_d=-0.3464$ and ${\bar a}_d=-0.5012$
for the vector and axial couplings of the first and second generation 
up-type and down-type quarks
taken from ref.~\cite{Field98}, we obtain $S_b=1.3184$. 
The $b$-quark contribution is
separated out in the quantity ${\bar s}_b$; here
$\mu_b$ is a correction factor coming from the nonzero $b$-quark mass.
This correction factor is roughly $\mu_b  \simeq 1.0 \times 10^{-3}$,
where we have taken
the running $b$-quark mass in the $\overline{\rm MS}$
scheme evaluated at $M_Z$, $m_b(M_Z) = 3.0$~GeV \cite{Fusaoka98}.

We can solve the above equations for ${\bar g}_b^L$ and ${\bar g}_b^R$ 
in terms of the experimentally measured values for $R_b$ and $A_b$.
Using the predicted SM values given in ref.~\cite{Field98}:
\begin{equation}
(\bar{g}^L_b)_{\rm SM} = -0.4208\,,  \qquad\qquad
(\bar{g}^R_b)_{\rm SM} = 0.0774\,,
\end{equation}
we obtain the SM predictions for $R_b$ and $A_b$:
\begin{eqnarray} 
                        R_b^{\rm SM} &=& 0.2158\,, \label{rbsm} \\
                        A_b^{\rm SM} &=& 0.935\,.  \label{absm}
\end{eqnarray}
These results should be compared with the measured values \cite{Mnich99}
\begin{eqnarray}
R_b &=& 0.21642 \pm 0.00073\,,
                        \label{eqn:Rbmeasured} \\
A_b &=& 0.893 \pm 0.016\,.
                        \label{eqn:Abmeasured}
\end{eqnarray}
$R_b$ is measured directly at LEP and SLD.
$A_b$ is measured directly at SLD from the left-right forward-backward 
asymmetry, and indirectly at LEP from 
the measured value of $A_e$ and
the forward-backward asymmetry 
$A_{FB}^{0,b} = \frac{3}{4} A_{e} A_{b}$. 
The $R_b$ measurement is $0.8\sigma$ above the SM prediction, and the 
$A_b$ measurement is $2.6\sigma$ below the SM prediction.

Allowing for a deviation of the experimentally measured values of
${\bar g}_b^{L,R}$ from their predicted values in the SM, we write:
\begin{equation} \label{deltanew}
\left(\bar{g}_b^{L,R}\right)_{\rm expt} =
      \left(\bar{g}_b^{L,R}\right)_{\rm SM} + \delta g^{L,R}_{\rm new}\,.
\end{equation}
The experimental constraints from $R_b$ and $A_b$ on 
$\delta g^{L,R}_{\rm new}$ are shown in fig.~\ref{fig:RbAb}.
\begin{figure}
\resizebox{\textwidth}{!}{\rotatebox{270}{\includegraphics{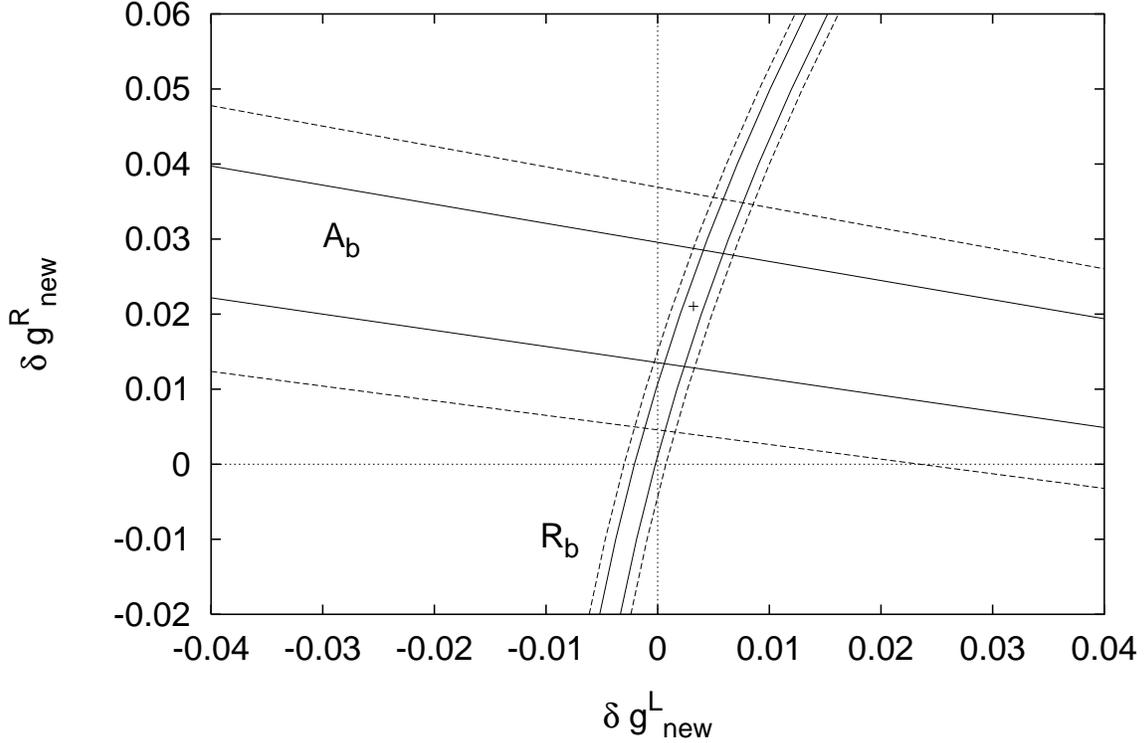}}}
\caption[Constraints on $\delta g^{R,L}$ from $R_b$ and $A_b$]
{The constraints from $R_b$ and $A_b$ on the right-- and   
left--handed $Z b \bar{b}$ couplings.  Plotted are the allowed deviations
$\delta g^{R,L}_{\rm new}$ of
the couplings from their SM values.  The 1$\sigma$
errors are shown as solid lines and the 2$\sigma$ errors as dashed lines.
The central value, at 
$\delta g^L_{\rm new} = 0.0037$ and 
$\delta g^R_{\rm new} = 0.0219$, is marked by the cross.
}
\label{fig:RbAb}
\end{figure}
The central value is at $\delta g^L_{\rm new} = 0.0037$ and $\delta
g^R_{\rm new} = 0.0219$.  Comparing these to the SM predictions, we
see that $\delta g^L_{\rm new}$ is roughly a $1\%$ correction while
$\delta g^R_{\rm new}$ is close to a $30\%$ correction.

It is also useful to expand $R_b$ and $A_b$ about their SM values, to
first order in $\delta g^{L,R}_{\rm new}$.  Using the SM parameters given 
above, we find
\begin{eqnarray}\label{Rbsigns}
\delta R_b &=& -0.7785\,\delta g^L_{\rm new} + 0.1409\,\delta g^R_{\rm new}  
\,,\nonumber \\
\delta A_b &=& -0.2984\,\delta g^L_{\rm new} - 1.6234\,\delta g^R_{\rm new}.
\end{eqnarray}
Note that a positive $\delta g^L_{\rm new}$ decreases both $R_b$ and $A_b$, 
while a positive $\delta g^R_{\rm new}$ increases $R_b$ and decreases $A_b$.
Inverting the above results yields
\begin{eqnarray}
\delta g^L_{\rm new} &=& -1.2433\, \delta R_b - 0.1079\, \delta A_b\,,
 \nonumber\\
\delta g^R_{\rm new} &=& 0.2286\, \delta R_b - 0.5962\, \delta A_b\,.
\end{eqnarray}
In practice, these first order results
provide a fairly good estimate of $\delta g^L_{\rm new}$, and a less reliable
estimate of $\delta g^R_{\rm new}$.  This is easily understood; because the
data suggest a rather large relative shift of ${\bar g}_b^R$ from its
SM predicted value, second order effects cannot be neglected.  In this
paper, the more precise analysis based on fig.~\ref{fig:RbAb} is used in
our analysis of new physics contributions to $R_b$ and $A_b$ from
extended Higgs sectors.\footnote{The bounds on Higgs sector parameters
obtained in section~\ref{sec:models} are based on a slightly older analysis of
electroweak data presented in ref.~\cite{Clare99}, which reported a
slightly higher value of $R_b$ and $A_b$. The effect of the updated
numbers on our plots is not significant and does not alter our general
conclusions.}

\subsection{Tree--level \boldmath$Zb\bar{b}$ couplings: The effect of oblique
corrections}
\label{sec:STU}

In the SM, all electroweak observables are fixed by the measurement of 
three quantities, commonly chosen to be the electromagnetic fine structure
constant $\alpha$, the muon decay constant $G_{\mu}$, and the $Z$ mass.
In particular, by measuring these quantities, one can predict the value of 
$\sin^2\theta^{\rm lept}_{\rm eff}$.  In practice, many more 
electroweak observables are measured and a fit is made to the SM parameters 
(see {\it e.g.}, ref.~\cite{CERN99}).

However, the dependence of $\sin^2\theta^{\rm lept}_{\rm eff}$ on other
electroweak observables can be modified in models of physics beyond
the SM.  The dominant effect of the new physics (in most cases) enters
via the virtual loop corrections to gauge boson self-energies; these are
the oblique corrections.  These modifications are parameterized
by the Peskin--Takeuchi parameters $S$, $T$, and $U$ \cite{Peskin92}.
In particular \cite{Grant98},
\begin{eqnarray} \label{deltasw}
\sin^2\theta^{\rm lept}_{\rm eff}-[\sin^2\theta^{\rm lept}_{\rm eff}]_{\rm SM}
                        \equiv \delta s^2_W
                       &=& \frac{\alpha}{c^2_W - s^2_W}
                        \left[ \nicefrac{1}{4} S - s^2_W c^2_W T
                       \right] \nonumber \\[6pt]
                       &=&3.40\times 10^{-3}\,S-2.42\times 10^{-3}\,T\,,
\end{eqnarray}
where we have used $s_W^2\equiv [\sin^2\theta^{\rm lept}_{\rm eff}]_{\rm SM} 
= 0.231$.
Nonzero values of the $S$ and $T$ parameters therefore modify
the prediction for the tree--level $Zb\bar{b}$ couplings 
$g_{Zb\bar{b}}^{L,R}$.

The $S$, $T$ and $U$ parameters are defined relative to a reference SM,
with a fixed Higgs mass. 
For $M_{h^0} = M_Z$, a fit of the electroweak data gives \cite{Caso98}
\begin{eqnarray}
S &=& -0.16 \pm 0.14\,,  \nonumber \\
T &=& -0.21 \pm 0.16\,,  \nonumber \\
U &=& 0.25 \pm 0.24\,.
\end{eqnarray}
This analysis has not yet been updated to account for the latest available
precision electroweak data. However, for our purposes, it is
sufficient to note is that the fitted absolute values of $S$ and $T$
are significantly less than ${\cal O}(1)$.

In order to understand the significance of oblique corrections of this 
size, we compute the corrections to the predictions for 
$R_b$ and $A_b$ due to $S$ and $T$ (there is no $U$ dependence).  
To first order in $\delta s^2_W$, eq.~(\ref{deltanew}) is modified to
\begin{equation} \label{deltanewprime}
\left(\bar{g}_b^{L,R}\right)_{\rm expt} =
      \left(\bar{g}_b^{L,R}\right)_{\rm SM} + \delta g^{L,R}_{\rm new}
      +\frac{1}{3} \delta s_W^2\,. 
\end{equation}
The last term is simply a consequence of the form of the $Zb\bar b$
tree-level couplings.  Since $A_b$ depends only on ${\bar g}_b^{L,R}$,
one may simply combine the results of eqs.~(\ref{Rbsigns}),
(\ref{deltasw}) and (\ref{deltanewprime}) to obtain:
\begin{equation}
\delta A_b = -0.641 \delta s^2_W
              = -2.18 \times 10^{-3}\, S  + 1.55 \times 10^{-3}\, T\,.
\end{equation}
To obtain $\delta R_b$, one must also account for the effect of the
oblique corrections on $g_u^{L,R}$ and $g_d^{L,R}$ which enter in the
expression for $\Gamma(Z\to~{\rm hadrons})$.  Following
ref.~\cite{Grant98}, we find:
\begin{equation}
\delta R_b =0.0388 \delta s^2_W
          = 1.32 \times 10^{-4}\, S  - 0.94 \times 10^{-4}\, T  \,.
\end{equation}
For values of $S$ and $T$ significantly less than ${\cal O}(1)$, 
the shift in the predicted value of $R_b$ and $A_b$ due to
nonzero values of $S$ and $T$ is less than a few percent of the present
experimental error on both $R_b$ and $A_b$.  We can therefore safely neglect
these corrections.

\section{Models with extended Higgs sectors}
\label{sec:extendedHiggs}

A wide variety of extensions to the minimal SM Higgs sector are 
possible \cite{HHG}.
We assume that the Higgs sector contains at least one complex
${\rm SU}(2)_L$ doublet with $Y= 1$ to give mass to the SM fermions.  
In our notation, $\phi_k$
denotes a multiplet of scalar fields that 
transforms as a complex representation under 
SU(2)$\times$U(1).\footnote{Given a complex Higgs multiplet, $\Phi$, with
$Y\neq 0$, one can always construct the complex conjugated multiplet,
$\Phi^*$, with hypercharge $-Y$.  Henceforth, without loss of
generality, we shall focus only on Higgs multiplets with $Y\geq 0$.}
A real representation ({\it i.e.}, a real multiplet of fields with 
integer weak isospin and hypercharge $Y=0$)
is denoted by $\eta_{i}$.  For simplicity, we assume that the Higgs
sector is CP--conserving, so that the neutral Higgs mass eigenstates are
either CP--even ($H^0_i$) or CP--odd ($A^0_j$).  The Higgs
potential is chosen to break ${\rm SU}(2)_L \times {\rm U}(1)_Y$ down to 
${\rm U}(1)_{\rm EM}$.  That is, we assume that only the neutral
member of each 
Higgs multiplet can acquire a non-zero vacuum expectation value (vev). 
For the neutral scalar component of a complex representation, the vev
is normalized such that
\begin{equation}
\phi_k^0\equiv\sqrt{\half}\left(v_k+\phi^{0,r}_k+i\phi^{0,i}_k\right)\,,
\end{equation}
where $\langle \phi^0_k \rangle = v_{k}/\sqrt{2}$.  For real
representations, we take $\langle \eta_i^0 \rangle = v_{i}$.  

Given the Higgs representations and the vevs, the Goldstone bosons 
eigenstates are determined.  The neutral Goldstone boson is given by
\begin{equation}
G^{0} = \left[\sum_{k} v_{k}^{2} Y_{k}^{2}\right]^{-1/2}
\sum_{k} v_k^2 Y_{k}^2\phi_{k}^{0,i}\,,
\end{equation}
and the positively charged Goldstone boson is given by
\begin{eqnarray} \label{eq:G+}
&& G^{+} =  N^{-1}\left[
            \sum_{k}\left\{ \left[T_{k}(T_{k}+1) - \quarter Y_{k}
                (Y_{k}-2) \right]^{1/2} v_{k}\phi_{k}^{+} 
                                  \right. \right. \nonumber \\
          && \,\,         - \left. \left.  \left[T_{k}(T_{k}+1)
     - \quarter Y_{k}(Y_{k}+2) \right]^{1/2} v_{k}(\phi_{k}^{-})^{*} \right\}
         + \sum_{i} \left[ 2T_{i}(T_{i}+1) \right]^{1/2} v_{i}\eta_{i}^{+} 
                                  \right] ,
\end{eqnarray}
where the normalization factor is given by
\begin{equation}  \label{nfactor} 
  N^2 \equiv \sum_{k}2v_{k}^{2}\left[T_{k}(T_{k}+1)-\quarter Y_{k}^{2}\right]
                        + \sum_{i}2v_{i}^{2}T_{i}(T_{i}+1) \,.
\end{equation}

In the above equations, we have separated out the sums into
contributions from the complex Higgs representations $k$ and the real
Higgs representations $i$.
Note that for a Higgs boson in a complex representation, $(\phi^Q)^*$ is a 
state with charge $-Q$ but is not the same as $\phi^{-Q}$.
For a Higgs boson in a real representation, we adopt the phase convention 
such that $(\eta^+)^* = - \eta^-$.  Thus, in our phase convention,
the negatively charged Goldstone boson is given by $G^- = - (G^+)^*$

Since we wish to preserve U(1)$_{\rm EM}$, we assume that only neutral
Higgs fields acquire vevs.  These Higgs vevs are constrained by the $W$ 
mass, which for a general extended Higgs sector is given by
\begin{equation} \label{wmass}
M_W^2=\quarter g^2 N^2 = \quarter g^2 v_{\rm SM}^2\,,
\end{equation}
where $N^2$ is given by eq.~(\ref{nfactor}).  Thus, we can identify
$N\equiv v_{\rm SM}= 246$~GeV. 

The vevs and/or the Higgs representation content
are also constrained by the $\rho$-parameter, which at tree-level 
is given by \cite{HHG}
\begin{equation} \label{rhohiggs}
\rho \equiv \frac{m_{W}^{2}}{M_{Z}^{2} c_{W}^{2}} = 
\frac{N^2}    {\sum_{k} v_k^2 Y_{k}^{2}}\,.
\end{equation}
The observed electroweak data imply that the tree-level value of
$\rho$ must be very close to (or perhaps exactly equal to) unity. 

In a Higgs sector that contains only multiplets which satisfy the
relation
\begin{equation}
(2 T + 1)^2 - 3 Y^2 = 1,
\label{eqn:rhois1}
\end{equation}
one finds $\rho = 1$ at tree level for any combination
of vevs.  Eq.~(\ref{eqn:rhois1}) is satisfied, for example, by the
familiar Higgs doublet with $Y= 1$, and by a series of larger
multiplets \cite{Tsao1}.\footnote{Of course, one can always add gauge neutral
singlet scalars with arbitrary vevs, without affecting the value of
the $\rho$-parameter.} 
In such a Higgs sector, the formulae for $G^+$ and $M_W^2$ simplify to
\begin{equation}
G^{+} = \left[\sum_{k} v_k^2 Y_{k}^{2}\right]^{-1/2}\,
\sum_{k}\sqrt{\half} v_k\left[(Y_k^2 + Y_k)^{1/2} \phi_{k}^{+}
        - (Y_k^2 - Y_k)^{1/2} (\phi_{k}^{-})^{*} \right]\,,
\end{equation}
and 
\begin{equation}
M_{W}^2 = \quarter g^2 \sum_{k} v_k^2 Y_{k}^{2}\,.
\end{equation}

In the SM, the diagonalization of the quark mass matrix automatically
diagonalizes the Yukawa couplings of the neutral Higgs boson to quarks.
Thus in the SM, there are no FCNC's mediated by tree--level 
Higgs exchange.  However, in a multi-doublet Higgs sector with the most 
general Higgs-fermion Yukawa couplings, tree-level Higgs-mediated FCNC's 
can arise.  These can be automatically
eliminated in any Higgs model in which fermions of a given electric charge
receive their mass from couplings to exactly one
neutral Higgs field \cite{Glashow77,Paschos77}.  This pattern of
Higgs-fermion couplings can be implemented by a judicious choice of
discrete symmetries.  There are two possible configurations for the 
Higgs-quark Yukawa couplings in an extended Higgs sector that contains at
least one scalar doublet with $Y= 1$.
In Type I models, all the quarks couple to one doublet, $\Phi_1$.  
In Type II models, the 
down--type quarks couple to $\Phi_1$ and the up--type quarks
couple to a second $Y= 1$ doublet, $\Phi_2$.  
If the general extended Higgs sector
contains only one $Y= 1$ doublet, then its Yukawa couplings are necessarily
Type~I.

In a Type I model, one Higgs doublet $\Phi_{1}$ gives mass to both 
$t$ and $b$ quarks.  The Yukawa couplings are
\begin{equation} \label{eqn:lambdaI}
\lambda_t = \frac{\sqrt{2} m_t}{v_1}\,,    \qquad\qquad
\lambda_b = \frac{\sqrt{2} m_b}{v_1}\,.
\end{equation}
Note that in a Type I model, $\lambda_b / \lambda_t = m_b/m_t$,
so $\lambda_b \ll \lambda_t$ for all values of $v_1$.

In a Type II model, 
$\Phi_{1}$ couples to $b$ quarks and 
$\Phi_{2}$ couples to $t$ quarks.  The quark Yukawa couplings are then
\begin{equation} \label{eqn:lambdaII}
\lambda_t = \frac{\sqrt{2} m_t}{v_2}\,,\qquad\qquad
\lambda_b = \frac{\sqrt{2} m_b}{v_1}\,.
\end{equation}
Note that in a Type II model, $\lambda_b/ \lambda_t = (m_b/m_t) (v_2/v_1)$,
so $\lambda_b$ can be enhanced relative to $\lambda_t$ 
by choosing $v_1 \ll v_2$.

When the Higgs mass--squared matrix is diagonalized, 
the electroweak eigenstates 
mix to form mass eigenstates.  
The couplings of the Higgs mass eigenstates to quarks take the form
\begin{equation}
i \left(g^{L}_{H\bar{q}q} P_{L} + g^{R}_{H\bar{q}q} P_{R}\right)
      = i \left(g^{V}_{H\bar{q}q} + g^{A}_{H\bar{q}q} \gamma_{5}\right).
\end{equation}
The individual couplings to $b\bar{b}$ and $b\bar{t}$ in a Type II
model are given by
\begin{eqnarray}
g_{H_i^0b\bar b}^V &=& - \frac{\lambda_b}{\sqrt 2} 
              \langle H_{i}^{0} | \phi_{1}^{0,r} \rangle\,, \label{eq:gV}\\
%
g_{A_i^0b\bar b}^A &=& - \frac{i\lambda_b}{\sqrt 2}
        \langle A_{i}^{0} | \phi_{1}^{0,i} \rangle\,,   \label{eq:gA}\\
%
g_{H_i^+\bar t b}^R &=& - \lambda_b
             \langle H_{i}^{+} | \phi_{1}^{+} \rangle\,,   \label{eq:gR}\\
%
g_{H_i^+\bar t b}^L &=& + \lambda_t
        \langle H_{i}^{+} | \phi_{2}^{+} \rangle\,,   \label{eq:gL}
\end{eqnarray}
where the bracket notation is used to indicate the overlap between the
corresponding mass-eigenstate and interaction-eigenstate.
The Type I model couplings are obtained by 
replacing $\phi_2^+$ with $\phi_1^+$ in eq.~(\ref{eq:gL}); the
other couplings remain the same.

The $Z$--Higgs--Higgs couplings take the form 
$i g\ls{ZH_1H_2} (p_1 - p_2)^{\mu}$, where $p_1$ [$p_2$] is the incoming
momentum of $H_1$ [$H_2$].
The $Z$--Higgs--Higgs couplings involving neutral and 
singly--charged Higgs bosons are
\begin{eqnarray}
g\ls{ZH_i^0A_j^0} &=& \frac{ie}{s_Wc_W} \sum_{k=1}^N
                     \langle H_{i}^{0} | \phi_{k}^{0,r} \rangle
                     \langle A_{j}^{0} | \phi_{k}^{0,i} \rangle
                     T^{3}_{\phi_{k}^{0}}\,,       \label{eq:ZHA} \\
g\ls{ZH_i^+H_j^-} &=& - \frac{e}{s_Wc_W} \left\{ \sum_{k=1}^N
                \langle H_{i}^{+} | \phi_{k}^{+} \rangle
                \langle H_{j}^{+} | \phi_{k}^{+} \rangle T^{3}_{\phi_{k}^{+}}
                        - s_{W}^{2} \delta_{ij}  \right\}\,,
\label{eq:Z+-}
\end{eqnarray}
where $T^{3}_{\phi}$ is the third component of the weak isospin of 
$\phi$. 
For completeness, we also give the $W^+W^-H_i^0$ and $ZZH_i^0$ couplings,
which take the form $i g\ls{V_1 V_2H} g^{\mu\nu}$.  
The $VVH_i^0$  ($V=W^\pm$, $Z$) couplings are
\begin{eqnarray}
g\ls{W^+W^-H_i^0}
&=& g^{2} \sum_{k}\langle H_i^{0}|\phi_{k}^{0,r} \rangle v_{k}
\left[T_{k}(T_{k}+1) - \quarter Y_{k}^{2}\right]\,,
\label{eq:WWH} \\
g\ls{ZZH_i^0} &=& \frac{g^2}{2c^2_W} \sum_k
                 \langle H_i^{0}|\phi_{k}^{0,r} \rangle v_{k} Y^2_k\,.
\label{eq:ZZH}
\end{eqnarray}
A complete list of Higgs--vector boson couplings in 
a general extended Higgs sector can be found in 
ref.~\cite{hlthesis}.

Although the $Z$--Higgs--Higgs couplings are diagonal in the interaction
basis, they are not necessarily diagonal in the mass-eigenstate basis.
In addition, the $ZH^{+}H^{-}$ couplings can differ from the SM
$ZG^{+}G^{-}$ coupling.  This can happen in a general model
if $H^+$ has some admixture of a multiplet larger than a doublet.
In the SM, the $ZG^+G^-$ coupling is
\begin{equation}
g\ls{ZG^+G^-} =  - \frac{e}{s_Wc_W}
                        \left(\nicefrac{1}{2} - s_{W}^{2} \right)\,.
\end{equation}

\section{Charged Higgs corrections to \boldmath$Z \rightarrow b \bar b$}
\label{sec-H+}

In the SM, the $Zb\bar b$ couplings
receive a correction from the exchange of the longitudinal components
of the $W^{\pm}$ and $Z$ bosons.  The
Feynman diagrams for these corrections are shown in 
fig.~\ref{fig:gloops}.
\begin{figure}
\begin{center}
\resizebox{8cm}{!}
        {\includegraphics*[100pt,430pt][370pt,700pt]{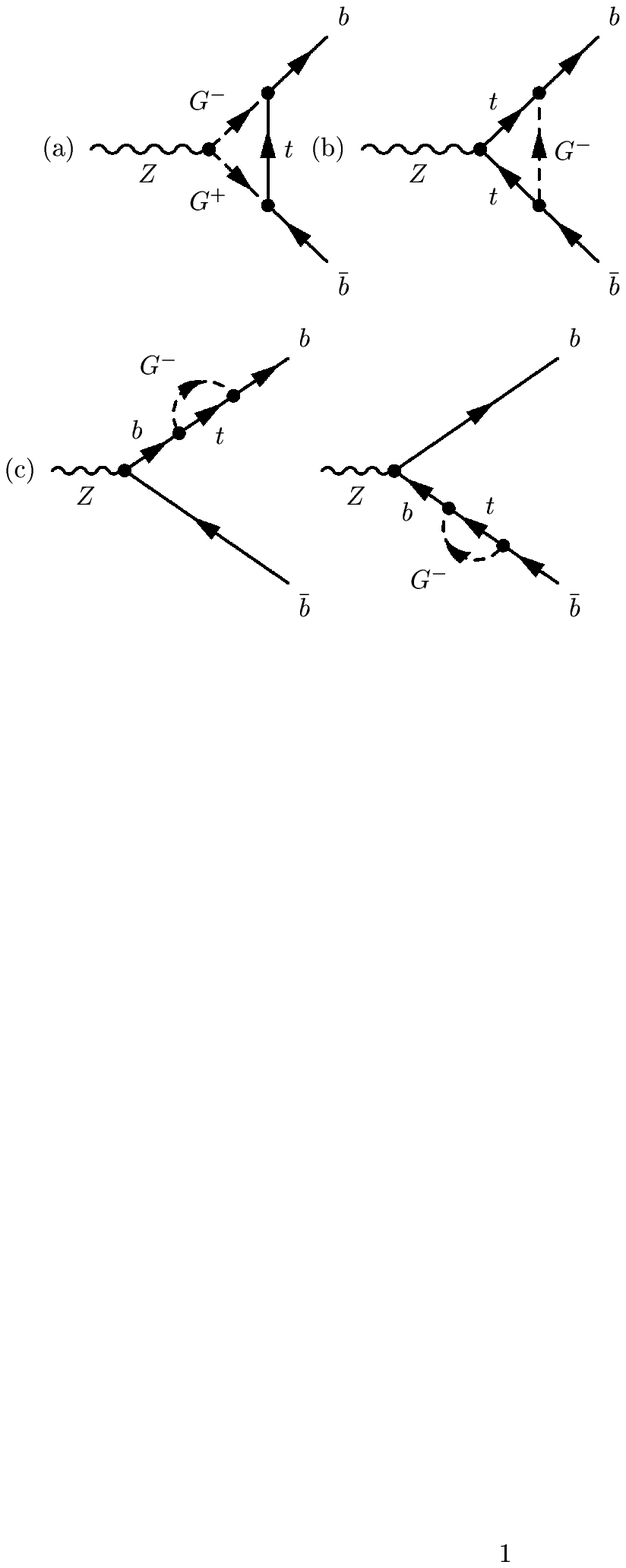}}
\end{center}
\caption
{Feynman diagrams of the leading $m_t^2$ contributions to the 
electroweak corrections to
$Z\rightarrow b \bar b$ in the SM.}
\label{fig:gloops}
\end{figure}
                        %
                        %
We work in the 't Hooft-Feynman gauge, in which the longitudinal
components of $W^{\pm}$ and $Z$ are just the Goldstone bosons
$G^{\pm}$ and $G^0$.  
The diagrams in fig.~\ref{fig:gloops} yield the leading 
$m_t^2$ contribution to $\delta g^{L,R}$ in the SM.  
A detailed review of the calculation of these diagrams is given in
ref.~\cite{Hollik2}.  Six additional diagrams, where one or two of the
$G^\pm$ lines in fig.~\ref{fig:gloops} is replaced by a 
corresponding $W^\pm$ line, also contribute to $\delta g^{L,R}$.
However, the latter
contributions are suppressed by a factor of $M_Z^2/m_t^2$ compared 
to the diagrams of fig.~\ref{fig:gloops}.

In an extended Higgs sector which contains singly charged Higgs states
$H_i^{\pm}$, the corrections to $\delta g^{L,R}$
arise from the diagrams of fig.~\ref{fig:h+loops}, where
$H_i^{\pm}$ runs over all the singly charged states in the Higgs sector,
including $G^{\pm}$.

\begin{figure}
\begin{center}
\resizebox{8cm}{!}
        {\includegraphics*[100pt,295pt][370pt,700pt]{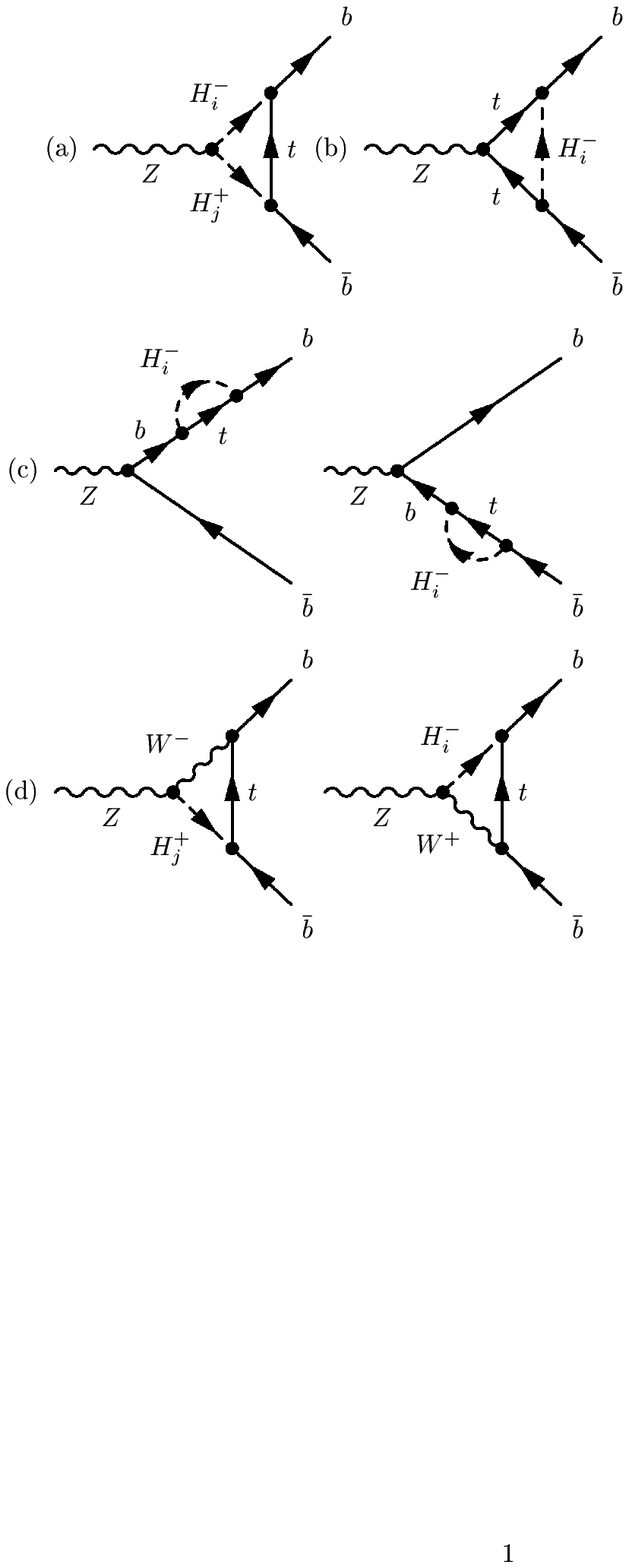}}
\end{center}
\caption
{Feynman diagrams of the electroweak corrections to
$Z\rightarrow b \bar b$ in a model
with an extended Higgs sector. 
}
\label{fig:h+loops}
\end{figure}

In calculating the corrections shown in fig.~\ref{fig:h+loops}
we keep only the leading term
in powers of $m_t^2 / M_Z^2$.  In $\delta g^L$ this leading term 
is proportional to $m_t^2$, where the two powers of $m_t$ come from
the left--handed Higgs--quark couplings $g^L_{H^+_i \bar t b}$.
In $\delta g^R$ the right--handed Higgs--quark couplings are proportional
to $m_b^2 \tan^2\beta$, so the leading term in $\delta g^R$ 
does not grow with increasing $m_t$.
This approximation has been used in calculating the
leading $m_t^2$
corrections to $R_b$ in the SM in the classic papers
\cite{Beenakker88,Bernabeu88,Hollik90,Lynn90}, 
and in calculating the corrections in extended Higgs sectors in 
refs~\cite{Kundu96,Denner91,Djouadi91,Boulware91,Grant}.

The two diagrams in fig.~\ref{fig:h+loops}(d) 
involving a $ZW^+H^-_i$ vertex can be nonzero in 
models containing Higgs multiplets 
larger than doublets.  However, their contribution to $R_b$ and $A_b$
is suppressed by a factor of $M_Z^2/m_t^2$ compared to diagrams 
\ref{fig:h+loops}(a), (b)
and (c), and we will neglect them.
Diagrams \ref{fig:h+loops}(a), (b), and (c) yield
                        %
\begin{eqnarray}
\delta g^{L,R}(a) &=& \frac{1}{8 \pi^2} \sum_{i,j} 
g^{L,R}_{H^+_i\bar tb} g^{L,R}_{H^+_j\bar tb} 
g_{ZH^+_iH^{-}_j} 
C_{24}(m_b^2, M_Z^2, m_b^2; m_t^2, M_i^2, M_j^2)\,,
\nonumber \\
\delta g^{L,R}(b) &=& - \frac{1}{16 \pi^2} \sum_i
(g^{L,R}_{H^+_i\bar tb})^2 \Bigl\{\nicefrac{1}{2}g^{R,L}_{Zt\bar t}
 \nonumber \\
&&\qquad +
\left[-2g^{R,L}_{Zt\bar t}C_{24} + g^{L,R}_{Zt\bar t}m_t^2C_{0}\right]
(m_b^2, M_Z^2, m_b^2; M_i^2,m_t^2,m_t^2)\Bigr\}\,, 
\nonumber \\
\delta g^{L,R}(c) &=& \frac{1}{16 \pi^2} \sum_i
\left(g^{L,R}_{H^+_i\bar tb}\right)^2 g^{L,R}_{Zb\bar b}
B_{1}(m_b^2; m_t^2, M_i^2)\,.
\end{eqnarray}
For the two-- and three--point integrals $C_{24}$, $C_{0}$, and 
$B_{1}$, we follow the definitions and conventions of ref.~\cite{Hollik1}.
The sums over $i$ and $j$ run over all the singly
charged Higgs mass eigenstates $H_i^+$ as well as the Goldstone 
boson $G^+$.  Where no ambiguity is involved, we have given the 
arguments of groups of tensor integrals that depend on the same 
variables only once at the end of the group.  These expressions for 
$\delta g^{L}$ agree with those of ref.~\cite{Kundu96}.

Collecting the results, and expressing the corrections in terms 
of the quark Yukawa couplings, we obtain for a Type II model
\begin{eqnarray}
\label{eqn:dgLH+}
\delta g^L &=& -\frac{\lambda_t^2}{16 \pi^2}  \frac{e}{s_Wc_W}
   \sum_{i,j}
   \vev{H^+_i | \phi^+_2} \vev{H^+_j | \phi^+_2} \nonumber \\
& & \qquad\times  \left\{ \sum_{k=1}^N
   \vev{H_i^+ | \phi_k^+}
   \vev{H_j^+ | \phi_k^+} T^3_{\phi_k^+}
   - s_W^2 \delta_{ij}  \right\}  
    2 C_{24} (m_t^2, M_i^2, M_j^2)\nonumber \\
& & - \frac{\lambda_t^2}{16 \pi^2}  \sum_i 
     \vev{H^+_i | \phi^+_2}^2 \Bigl\{\nicefrac{1}{2}g^R_{Zt\bar t} +
     \left[-2 g^R_{Zt\bar t} C_{24}
 + g^L_{Zt\bar t} m_t^2 C_0 \right] (M_i^2, m_t^2, m_t^2)\Bigr\}  \nonumber \\
& & + \frac{\lambda_t^2}{16 \pi^2} g^L_{Zb\bar b} \sum_i 
     \vev{H^+_i | \phi^+_2}^2  B_1 (m_b^2; m_t^2, M_i^2)\,,
\end{eqnarray}
\begin{eqnarray}
\label{eqn:dgRH+}
\delta g^R  &=& -\frac{\lambda_b^2}{16 \pi^2}  \frac{e}{s_Wc_W}
     \sum_{i,j}
     \vev{H^+_i | \phi^+_1} \vev{H^+_j | \phi^+_1} \nonumber \\
& & \qquad \times\left\{\sum_{k=1}^N
     \vev{H_i^+ | \phi_k^+}
     \vev{H_j^+ | \phi_k^+} T^3_{\phi_k^+}
      - s_W^2 \delta_{ij}  \right\}
      2 C_{24} (m_t^2, M_i^2, M_j^2)  \nonumber \\
& & - \frac{\lambda_b^2}{16 \pi^2} \sum_i 
      \vev{H^+_i | \phi^+_1}^2 \Bigl\{\nicefrac{1}{2}g^L_{Zt\bar t} +
      \left[-2 g^L_{Zt\bar t} C_{24} 
  + g^R_{Zt\bar t} m_t^2 C_0 \right] (M_i^2, m_t^2, m_t^2)\Bigr\}  \nonumber \\
& & + \frac{\lambda_b^2}{16 \pi^2} g^R_{Zb\bar b} \sum_i
      \vev{H^+_i | \phi^+_1}^2   B_1 (m_b^2; m_t^2, M_i^2)\,.
\end{eqnarray}
For compactness we have dropped
the first three arguments of the three--point integrals, 
$(m_b^2, M_Z^2, m_b^2)$, because these arguments are the same in all the
expressions.  The first three arguments of the three--point integrals
depend only on the masses of the on--shell external particles.

The corrections for a Type I model are obtained by replacing 
$\phi_2^+$ with $\phi_1^+$ in $\delta g^L$.
We see that $\delta g^L$ is proportional to $\lambda_t^2$ and 
$\delta g^R$ is proportional to $\lambda_b^2$.  Clearly,
$\delta g^R$ is negligible 
compared to $\delta g^L$, except in a Type II model when $\lambda_b$ 
is enhanced for small $v_1$.
In this situation there is also a significant contribution to 
$\delta g^{L,R}$ coming from loops involving the neutral
Higgs bosons, as described in the next section.

In the Type II 2HDM, $\delta g^R$ is proportional to 
$(m_b \tan \beta)^2$, while 
$\delta g^L$ is proportional to $(m_t \cot \beta)^2$. 
At large $\tan\beta$,
$\delta g^R$ is enhanced and $\delta g^L$
is suppressed; $\lambda_t$ and $\lambda_b$ are the same size
when $\tan\beta = m_t/m_b \simeq 50$.
However, because of their different dependence on the 
$Zq\bar{q}$ couplings,
$\delta g^L$ and $\delta g^R$ are the same size 
when $\tan \beta \simeq 10$. 

The formulae in eqs.~(\ref{eqn:dgLH+})--(\ref{eqn:dgRH+})
can be simplified.  Electromagnetic gauge invariance requires
that the terms proportional to $s^2_W$ (from the $Zq\bar{q}$ and 
$ZH^+H^{-}$ couplings) add to zero in the limit $M_Z^2 \rightarrow 0$.  
This provides a check of our calculations.  In our approximation
we neglect terms of order $M_Z^2/m_t^2$.
Using the expansions for
the two-- and three--point integrals given in ref.~\cite{PHI}
and neglecting terms of order 
$M_Z^2/m_t^2$ in the three--point integrals, 
we find that the terms proportional to 
$s^2_W$ cancel.  The corrections can then be written as
%
\begin{eqnarray}
\delta g^{L,R} &=&  \mp \frac{1}{16\pi^2} \frac{e}{s_Wc_W}
       \sum_i
       \left(g_{H^+_i\bar tb}^{L,R}\right)^2
       \nicefrac{1}{2} m_t^2 C_{0}(M_i^2,m_t^2,m_t^2)
   \nonumber \\
 && - \frac{1}{16\pi^2} \frac{e}{s_Wc_W}
        \sum_i
       \left(g_{H^+_i\bar tb}^{L,R}\right)^2
       \sum_k \vev{H_i^+ | \phi_k^+}^2
       (T^3_{\phi_k^+} - \nicefrac{1}{2})
       2 C_{24}(m_t^2,M_i^2,M_i^2)
  \nonumber \\
 && - \frac{1}{16\pi^2} \frac{e}{s_Wc_W}
        \sum_i \sum_{j \neq i}
    \left(g_{H^+_i\bar tb}^{L,R}\right)
    \left(g_{H^+_j\bar tb}^{L,R}\right)
        \sum_k
        \vev{H_i^+ | \phi_k^+}
        \vev{H_j^+ | \phi_k^+} T^3_{\phi_k^+}
  \nonumber \\
 & & \qquad\qquad\qquad\qquad\times 2 C_{24}(m_t^2,M_i^2,M_j^2).
\label{eq:h+loops2}
\end{eqnarray}
The third term in eq.~(\ref{eq:h+loops2}) is the sum of the diagrams
\ref{fig:h+loops}(a) for two different charged Higgs bosons $H^+_i$ and
$H^+_j$ in the loop.  It is only nonzero when there are
nonzero off--diagonal $ZH^+_iH^-_j$ couplings ($i\neq j$).
The second term describes the contribution to diagrams \ref{fig:h+loops}(a)
from diagonal $ZH^+_iH^{-}_i$ couplings when
$T^3_{\phi_k^+}$ is different from $1/2$.  
This term is only nonzero when
the Higgs sector contains multiplets larger than doublets.
The first term comes from the sum of diagrams \ref{fig:h+loops}(b) and (c),
plus the remaining part of diagram \ref{fig:h+loops}(a) with 
$T^3_{\phi_k^+} = 1/2$.  This part of diagram
\ref{fig:h+loops}(a) is what we would get if we replaced all of the 
$ZH^+H^{-}$ couplings with the SM $ZG^+G^{-}$ coupling. 
Note that for $m_t \gg M_Z$, $C_0(M_i^2,m_t^2,m_t^2)$ is negative.
Therefore the first term of $\delta g^{L}$ ($\delta g^{R}$) is always
positive (negative) definite, which decreases the prediction for $R_b$.

From eq.~(\ref{eq:h+loops2}), one can deduce a number of results. 
First, if the Higgs sector contains only doublets and
singlets, $T^3_{\phi_k^+} = 1/2$ and there are no 
off-diagonal $ZH^+H^{-}$ couplings.  Then the second and third terms 
of eq.~(\ref{eq:h+loops2}) are
zero.  We are left with the first term
\begin{eqnarray}
\label{eqn:H+doubletsonly}
\delta g^{L,R} & = & \mp \frac{1}{16\pi^2}\  \frac{e}{2s_Wc_W}
           \sum_i
           \left(g_{H^+_i\bar tb}^{L,R}\right)^2 m_t^2
           C_{0}(M_i^2,m_t^2,m_t^2)    \nonumber \\
& = & \delta g^{L,R}_{\rm SM} \pm \frac{1}{16\pi^2}\ \frac{e}{2s_Wc_W}
           \sum_{i \neq G^+}
           \left(g_{H^+_i\bar tb}^{L,R}\right)^2
           \left[ \frac{R_i}{R_i-1} - \frac{R_i \log R_i}
               {(R_i-1)^2} \right]\,,
\end{eqnarray}
where $R_i \equiv m_t^2/M_i^2$.  The correction in the 
SM due to $G^{\pm}$ exchange is denoted by $\delta g^{L,R}_{\rm SM}$.
The non--SM piece of 
$\delta g^{L}$ [$\delta g^{R}$] is positive [negative] definite, 
both of which decrease $R_b$.
Therefore, in order for it to be possible to increase
$R_b$ through
charged Higgs boson loops, we must have a Higgs sector that contains 
multiplets larger than doublets.

Second, if all the $H_i^+$ are degenerate with $G^+$, we
can sum over the complete sets of states in the second and third 
terms of eq.~(\ref{eq:h+loops2}).  These terms cancel
and again we are left with
\begin{eqnarray}
\delta g^{L} &=&  \frac{\lambda_t^2}{16 \pi^2}\
         \frac{e}{2s_Wc_W} \left[ \frac{R}{R-1} - \frac{R \log R}
         {(R-1)^2} \right]\,,\nonumber  \\
\delta g^R &=& - \frac{\lambda_b^2}{16 \pi^2}\
         \frac{e}{2s_Wc_W} \left[ \frac{R}{R-1} - \frac{R \log R}
         {(R-1)^2} \right]\,,
\end{eqnarray}
with $R\equiv m_t^2 / M_W^2$.  This formula includes the SM correction
$\delta g^{L,R}_{\rm SM}$.  
As above, the non--SM piece of $\delta g^{L}$ [$\delta g^{R}$] is 
positive [negative] definite, both of which decrease $R_b$. 

In a Higgs sector that contains only
multiplets for which $\rho = 1$ automatically 
[eq.~(\ref{eqn:rhois1})],
the Goldstone boson
does not contribute to the second and third terms of 
eq.~(\ref{eq:h+loops2}) because there are no off--diagonal
$Z G^+ H_i^-$ couplings, and the $Z G^+ G^-$ coupling is the same 
as in the SM.
Thus in such a model, if all the $H_i^+$
(excluding $G^+$) are degenerate with mass $M$, we can again sum
over the complete sets of states in
the second and third terms of eq.~(\ref{eq:h+loops2}).  These
terms again cancel and we are left with
\begin{eqnarray}
\delta g^L &=& \delta g^L_{\rm SM}
        + \frac{\lambda_t^2}{16 \pi^2} 
        \left( 1 - \frac{v_2^2}{v_{\rm SM}^2} \right)
      \frac{e}{2s_Wc_W} \left[ \frac{R}{R-1} - \frac{R \log R}
            {(R-1)^2} \right]\,, \label{eqn:dgLdegenH+} \\
\delta g^R &=& \delta g^R_{\rm SM}
         - \frac{\lambda_b^2}{16 \pi^2} 
         \left( 1 - \frac{v_1^2}{v_{\rm SM}^2} \right)
         \frac{e}{2s_Wc_W} \left[ \frac{R}{R-1} - \frac{R \log R}
          {(R-1)^2} \right]\,,
\label{eqn:dgRdegenH+}
\end{eqnarray}
with $R\equiv m_t^2 / M^2$, for a Type II model.  The correction in a 
Type I model is obtained by replacing
$v_2$ with $v_1$ in eq.~(\ref{eqn:dgLdegenH+}).  As above, the non--SM
piece of $\delta g^{L}$ [$\delta g^{R}$] is positive [negative] definite,
both of which decrease $R_b$.

\section{Neutral Higgs corrections to \boldmath$Z \rightarrow b \bar b$}
\label{sec-H0}

The corrections to $Z \rightarrow b \bar b$ from neutral
Higgs boson loops, shown in fig.~\ref{fig:h0loops}, are proportional
to $\lambda_b^2$.  
\begin{figure}
\begin{center}
\resizebox{8cm}{!}
        {\includegraphics*[100pt,160pt][370pt,700pt]{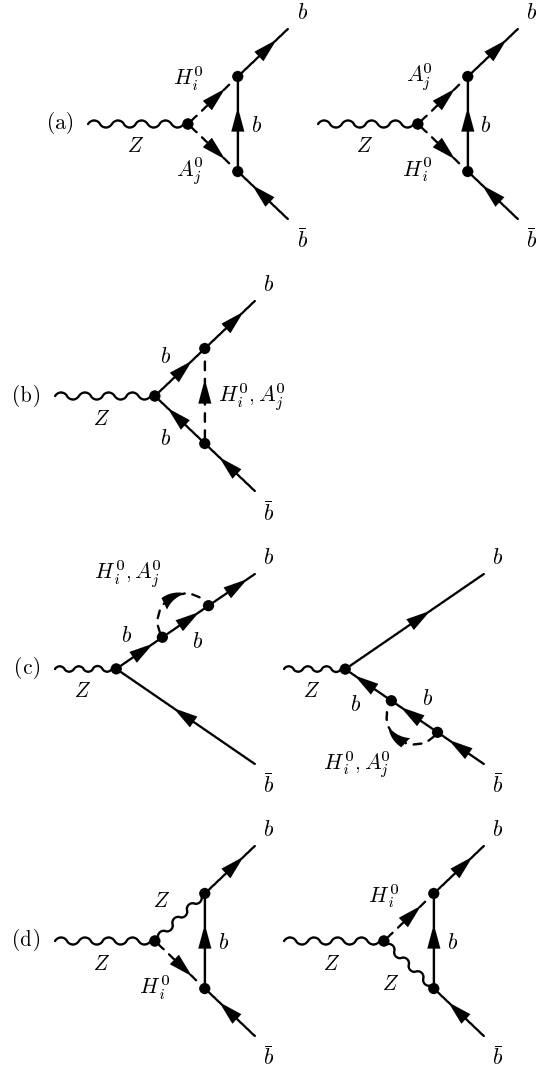}}
\end{center}
\caption{Feynman diagrams for the corrections to $Z \rightarrow b \bar b$
involving neutral Higgs bosons in the loop. 
}
\label{fig:h0loops}
\end{figure}
                        %
In a Type I
model, $\lambda_b \ll \lambda_t$, so the one-loop radiative corrections
mediated by
neutral Higgs bosons are negligible compared to charged Higgs mediated
corrections (which
are proportional to $\lambda_t^2$).  However, in
a Type II model, $\lambda_b$ increases as $v_1$ decreases.
In the limit of small $v_1$, the corrections mediated by neutral
Higgs bosons are significant.

In calculating the corrections due to the diagrams in fig.~\ref{fig:h0loops},
we neglect terms proportional to $m_b$ that are not enhanced by small $v_1$.
The diagrams of fig.~\ref{fig:h0loops}(d)
are suppressed by a factor of $m_b/M_Z$ 
compared to diagrams \ref{fig:h0loops}(a), (b) and (c), 
and so we neglect them as well.
The contributions to $\delta g^{R,L}$ from diagrams 
\ref{fig:h0loops}(a), (b), and (c) are
\begin{eqalignno}
\delta g^{R,L}(a)   
&= \pm \frac{1}{4\pi^2} \sum_{H^0_i,A^0_j}
          g\ls{ZH^0_iA^0_j} g^V_{H^0_ib\bar b}
          g^A_{A^0_jb\bar b}
          C_{24}(m_b^2,M_i^2,M_j^2)        \nonumber           \\
&= \mp \frac{\lambda_b^2}{8\pi^2}  \frac{e}{s_W c_W}
                        \sum_{H^0_i,A^0_j}
           \vev{H^0_i | \phi_1^{0,r} }
           \vev{A^0_j | \phi_1^{0,i} }                
     \sum_{k=1}^{N}
   \vev{H^0_i | \phi_k^{0,r} } \vev{A^0_j | \phi_k^{0,i} }
   T^3_{\phi_k^0} C_{24} (m_b^2, M_i^2, M_j^2)\,, \nonumber \\[6pt]
                        %
\delta g^{R,L}(b)   
&=         - \frac{1}{16\pi^2}\ g^{L,R}_{Zb\bar b}
                         \left[  
  \sum_{H^0_i} (g^V_{H^0_ib\bar b})^2
     \Bigl\{\nicefrac{1}{2} -\left[2 C_{24} + 
                M_{Z}^2 (C_{22}-C_{23}) \right]
              (M_i^2,m_b^2, m_b^2) \Bigr\} \right.         \nonumber \\
&\qquad\qquad
       \left. - \sum_{A^0_j} (g^A_{A^0_jb\bar b})^2
       \Bigl\{\nicefrac{1}{2}  -\left[2 C_{24} + 
        M_{Z}^2 (C_{22}-C_{23}) \right] (M_j^2,m_b^2, m_b^2)\Bigr\}
                                                 \right] \nonumber \\
&=
- \frac{\lambda_b^2}{32\pi^2} g^{L,R}_{Zb\bar b}
                        \left[
    \sum_{H^0_i} \vev{H^0_i | \phi_1^{0,r} }^2
    \Bigl\{\nicefrac{1}{2} -\left[2 C_{24} +M_{Z}^2 (C_{22}-C_{23})
    \right] (M_i^2, m_b^2, m_b^2) \Bigr\} \right.      \nonumber \\
&\qquad\qquad
\left.  +\sum_{A_j^0} \vev{A^0_j | \phi_1^{0,i} }^2
 \Bigl\{\nicefrac{1}{2} -\left[2 C_{24} + M_{Z}^2 (C_{22}-C_{23})
 \right] (M_j^2, m_b^2, m_b^2) \Bigr\}\right]\,,   \nonumber  \\[6pt] 
\delta g^{R,L}(c)    
&= \frac{1}{16\pi^2} g^{R,L}_{Zb\bar b} \!
          \left[
           \sum_{H^0_i} (g^V_{H^0_ib\bar b})^2
            B_{1}(m_b^2; m_b^2, M_i^2)      
         - \sum_{A^0_j} (g^A_{A^0_jb\bar b})^2
            B_{1}(m_b^2; m_b^2, M_j^2)
          \right]          \nonumber \\
&= \frac{\lambda_b^2}{32\pi^2} g^{R,L}_{Zb\bar b} 
    \left[  \sum_{H^0_i} \vev{H^0_i | \phi_1^{0,r} }^2 \!\!
            B_1 (m_b^2; m_b^2, M_i^2)        
   \!  +\! \sum_{A_j^0}\! \vev{A^0_j | \phi_1^{0,i} }^2 \!\!
      B_1 (m_b^2; m_b^2, M_j^2) \right]\,. 
\label{eq:h0loopsac}
\end{eqalignno}
For compactness of notation, we again drop the 
first three arguments, $(m_b^2, M_Z^2, m_b^2)$, 
of the three--point integrals.
Note that $g_{ZH_i^0A_j^0}$ and $g^A_{A_j^0bb}$ are imaginary,
while $g^V_{H_i^0bb}$ is real.  In the sums over scalar states, 
$H^0_i$ runs over all
CP-even neutral Higgs bosons, and $A^0_j$ runs over all CP--odd
neutral Higgs bosons (including $G^0$).  However, the corrections
involving $G^0$ can be neglected because the $G^0$ coupling
to $b\bar b$ is not enhanced by large $\lambda_b$.  
In particular,
$g^A_{G^0b\bar b} = - m_b / v_{\rm SM}$, independent of
the value of $v_1$.

As in section~\ref{sec-H+}, we can use electromagnetic gauge 
invariance to check our calculations.  Electromagnetic gauge
invariance requires that terms proportional to $s^2_W$ sum to
zero in the limit $M_Z \rightarrow 0$.  Note that $\delta g^{R,L}(a)$ is 
independent of $s^2_W$, whereas in the limit $M_Z \rightarrow 0$,
$\delta g^{R,L}(b) + \delta g^{R,L}(c) = 0$, independent of the
Higgs masses.  The terms proportional to $s^2_W$ indeed vanish
in this limit.  

Finally, we briefly 
examine the special case in which all the $H^0_i$ are degenerate
with mass $M_H$, and all the $A^0_j$ (excluding $G^0$) are degenerate with
mass $M_A$.  
In this case, we can sum over complete sets of states
and eq.~(\ref{eq:h0loopsac}) simplifies to
\begin{eqnarray}
\delta g^{R,L} (a) &=& \pm \frac{\lambda_b^2}{16 \pi^2}  
 \left( \frac{e}{s_W c_W} \right) C_{24} (m_b^2, M_H^2, M_A^2)\,,
\nonumber \\
\delta g^{R,L} (b) &=& - \frac{\lambda_b^2}{32 \pi^2} g^{L,R}_{Zb\bar b} 
  \Bigl\{1- \left[ 2 C_{24} +  M_Z^2 (C_{22} - C_{23})
 \right] (M_H^2, m_b^2, m_b^2)  \nonumber \\
 & & \qquad\qquad
-  \left[ 2 C_{24} +  M_Z^2 (C_{22} - C_{23})
 \right] (M_A^2, m_b^2, m_b^2) \Bigr\}\,,
\nonumber \\
\delta g^{R,L} (c) &=& \frac{\lambda_b^2}{32 \pi^2} g^{R,L}_{Zb\bar b} 
    \left[ B_1 (m_b^2; m_b^2, M_H^2) + B_1 (m_b^2; m_b^2, M_A^2) \right]\,.
\end{eqnarray}

\section{Corrections to \boldmath$Z \rightarrow b \bar{b}$ in specific 
extended Higgs models}
\label{sec:models}

In this section we calculate the radiative corrections to
$Z \rightarrow b \bar{b}$ in a variety of extended Higgs models,
and ascertain the constraints on the parameter space of
each model due to the experimental data.
%
%
We find that the corrections to $R_b$ are large enough that the measured
value 
of $R_b$ can be used to constrain the parameter space of specific models.
However, the corrections to $A_b$ are small compared to the uncertainty
in the measurement of $A_b$, and thus cannot be used to further
constrain the models.

\subsection{Models with Higgs doublets and singlets}
\label{sec:61nonexotic}

\subsubsection{Charged Higgs boson contributions}

In a model containing only Higgs 
doublets and singlets, the radiative corrections due to the charged 
Higgs bosons are described by eq.~(\ref{eqn:H+doubletsonly}).  
These corrections 
have definite signs; in particular, $\delta g^L > 0$ and $\delta g^R < 0$.  
Both of these give $\Delta R_b < 0$, in worse agreement with experiment 
than the SM.
The corrections due to neutral
Higgs boson exchange will also contribute when $\lambda_b$ is enhanced.
They must be taken into account as well in this regime when deriving
constraints from the $R_b$ measurement.

\paragraph{Two Higgs doublet model}

The 2HDM contains a single charged Higgs boson, 
\begin{equation}
H^+ = -\sin\beta\, \phi^+_1 + \cos\beta\, \phi^+_2.
\end{equation}
Its contribution to $\delta g^{L,R}$ is found from 
eq.~(\ref{eqn:H+doubletsonly}) with only one $H^+$ in the sum.
For the Type II 2HDM,
\begin{eqnarray}
\delta g^{L} &=& \frac{1}{32\pi^{2}} \left(\frac{gm_{t}}{\sqrt{2}M_{W}}
\cot \beta \right)^{2}
\frac{e}{s_{W}c_{W}} \left[ \frac{R}{R-1} - \frac{R \log R}
{(R-1)^{2}}  \right]\,,
\label{eq:dgl2HDM} \\
\delta g^{R} &=& -\frac{1}{32\pi^{2}} \left( \frac{gm_{b}}{\sqrt{2}M_{W}}
\tan \beta \right)^{2}
\frac{e}{s_{W}c_{W}} \left[ \frac{R}{R-1} - \frac{R \log R}
{(R-1)^{2}}  \right]\,,
\label{eq:dgr2HDM}
\end{eqnarray}
where $R\equiv m_{t}^{2}/M_{H^{+}}^{2}$.  This correction is in addition
to the correction due to Goldstone boson exchange, which is the same 
as in the SM.
This agrees with the results of 
refs.~\cite{Kundu96,Denner91,Djouadi91,Boulware91,Grant}.  
In the Type II
model, $\delta g^L$ is significant at small $\tan\beta$ and is suppressed
at large $\tan\beta$, while $\delta g^R$ is negligible at small $\tan\beta$
but is significant at large $\tan\beta$.

In a Type I model the result is the same except that $\cot^2\beta$ is
replaced with $\tan^2\beta$ in $\delta g^{L}$.  In this case, $\delta g^R$
is negligible compared to $\delta g^L$ at any value of $\tan\beta$.
Both $\delta g^L$ and $\delta g^R$ grow with increasing $\tan\beta$.


For small $\tan\beta$, the neutral Higgs couplings to $b$ quarks are
small, and contributions to $Z \rightarrow b \bar{b}$ due to 
neutral Higgs boson exchange can be neglected.  In this regime the
corrections due to charged Higgs boson exchange can be used to constrain
the 2HDM.
In fig.~\ref{fig:mtwohdm} we plot the constraints from $R_b$ on 
$M_{H^+}$ as a function of $\tan\beta$, for a Type II 2HDM.
                        %
We also show the constraints on
the charged Higgs mass from the process $b \to s \gamma$ 
\cite{CLEO99,Borzumati98} and the charged Higgs boson search
at LEP \cite{lephiggs}.
The constraint on the charged Higgs mass from the Tevatron D0 experiment
\cite{D0chargedH} is significantly weaker than the constraint 
from $b \to s \gamma$, and are not shown in fig.~\ref{fig:mtwohdm}.
$R_b$ provides the strongest constraint on $M_{H^+}$ for 
$\tan\beta < 1.5$.  For larger 
$\tan\beta$, the constraint from $b \rightarrow s \gamma$ 
is stronger.

\begin{figure}
\resizebox{\textwidth}{!}{\rotatebox{270}{\includegraphics{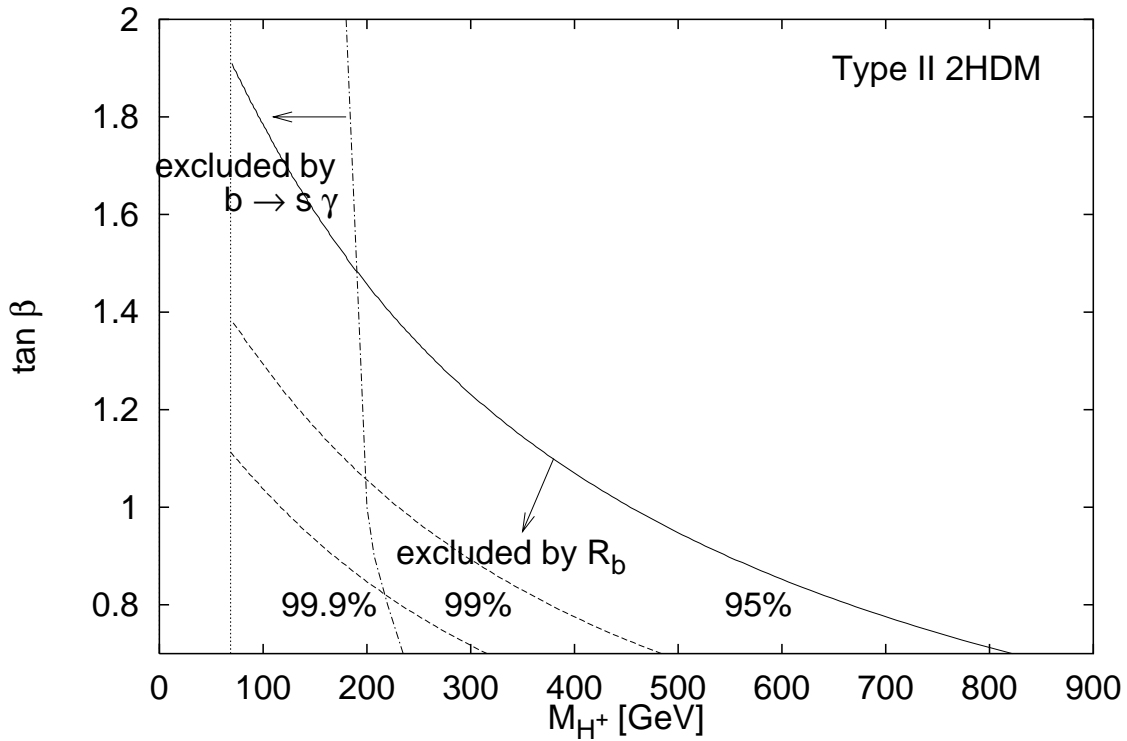}}}
\caption[Constraints from $R_b$ on the charged Higgs mass 
and $\tan\beta$ in the Type II 2HDM]
{Constraints from $R_b$ on the charged Higgs mass and $\tan\beta$ in the
Type II 2HDM.  The area below the solid line is excluded at 95\%
confidence level.  Also shown are
the 99\% and 99.9\% confidence levels (dashed lines).
We also show the 95\% confidence level lower bound on $M_{H^+}$
from the $b \rightarrow s \gamma$ branching ratio
\cite{CLEO99,Borzumati98}  (dot-dashed).
The vertical dotted line is the direct search bound on the charged
Higgs mass, $M_{H^+} > 77.3$~GeV \cite{lephiggs}.
}
\label{fig:mtwohdm}
\end{figure}

For large $\tan\beta$, neutral Higgs boson exchange contributes to 
$Z \rightarrow b \bar{b}$ in addition to charged Higgs boson exchange.  
The neutral Higgs boson contributions are discussed in 
section~\ref{sec:NeutralHiggs}.

In the case of a Type I 2HDM, the bound on $M_{H^+}$ from $R_{b}$ 
is the same as in fig.~\ref{fig:mtwohdm}, but with $\cot\beta$
replacing $\tan\beta$ on the vertical axis.  
In this class of models there is no constraint at present on the 
charged Higgs boson mass from $b \rightarrow s \gamma$.

\paragraph{Multiple--doublet models and models with singlets}

We now consider the effects of
charged Higgs boson exchange in a model containing multiple 
Higgs doublets, denoted $\Phi_k$, with $Y=1$.
We can add to this model any number of Higgs singlets 
with zero hypercharge.  These contain only neutral degrees of freedom,
and so they have no effect on the charged Higgs sector.

In a Type I model of this type, we let $\Phi_1$ couple to both up--
and down--type quarks, and none of the other doublets couple to quarks.
In a Type II model, we let $\Phi_1$ couple only to down--type quarks, 
and $\Phi_2$ couple to up--type quarks.  Then the Yukawa couplings 
are defined in the same way as in the 2HDM, in 
eqs.~(\ref{eqn:lambdaI})--(\ref{eqn:lambdaII}).

In a Type II model, the contributions to $Z \rightarrow b \bar{b}$ from
charged Higgs boson exchange are
\begin{eqnarray}
&&\delta g^L = \frac{1}{32 \pi^2} \frac{e}{s_W c_W} 
\left( \frac{gm_t}{\sqrt{2}M_W} \frac{v_{\rm SM}}{v_2} \right)^2
 \sum_{i \neq G^+} \langle H_i^+ | \phi_2^+ \rangle^2
 \left[ \frac{R_i}{R_i-1} - \frac{R_i \log R_i}{(R_i-1)^2} \right],
\label{eqn:dgLMHDM} \\                        %
&&\delta g^R = -\frac{1}{32 \pi^2} \frac{e}{s_W c_W} 
\left( \frac{gm_b}{\sqrt{2}M_W} \frac{v_{\rm SM}}{v_1} \right)^2
 \sum_{i \neq G^+} \langle H_i^+ | \phi_1^+ \rangle^2
 \left[ \frac{R_i}{R_i-1} - \frac{R_i \log R_i}{(R_i-1)^2} \right]
\end{eqnarray}
where $R_{i} \equiv m_{t}^{2}/M_{H_i^+}^{2}$.  
This contribution is in addition
to the contribution due to charged Goldstone boson exchange, which is the
same as in the SM.
In a Type I model, the contribution is the same except that $v_2$ is replaced
with $v_1$ and $\phi_2^+$ is replaced with $\phi_1^+$ in the formula 
for $\delta g^L$.

These corrections to $\delta g^{L,R}$ from charged Higgs boson exchange have
the same dependence on the charged Higgs masses as the corrections in
the 2HDM.  The contribution from each $H_i^+$ is weighted by the
overlap of each $H^+_i$ with the electroweak eigenstate
that couples to the quarks involved.

Note that the Yukawa couplings depend on the ratios $v_{\rm SM}/v_2$ 
and $v_{\rm SM}/v_1$.
This is the same dependence as in the 2HDM.  Recall that in the 2HDM, $v_1$ and
$v_2$ were constrained by the $W$ mass to satisfy the relation, 
$v_1^2 + v_2^2 = v_{\rm SM}^2$.
Thus in the 2HDM, $v_1$ and $v_2$ cannot both
be small at the same time.  However, in a model with more than two
doublets, the $W$ mass constraint involves the vevs of all the doublets
(labeled by $k$), giving $\sum_k v_k^2 = v_{\rm SM}^2$.
In this model, both $v_1$ and $v_2$ can be small at the
same time, leading to significant contributions to both $\delta g^L$ and
$\delta g^R$.

The corrections to $Z \rightarrow b \bar{b}$ in this model can be understood
by examining their behavior in certain limits.
First, let us examine the limit in which all but one of the $H_i^+$
are very heavy.  The contributions of the heavy $H_i^+$ to 
$\delta g^{L,R}$ go to zero as the masses go to infinity.  The remaining
contribution to $\delta g^{L,R}$ is due to the single light charged Higgs
boson,
and it is of the same form as in the 2HDM.  Comparing with 
eqs.~(\ref{eq:dgl2HDM})--(\ref{eq:dgr2HDM}), we see that in $\delta g^L$,
$\tan\beta$ is replaced by 
$v_2/[v_{\rm SM}\langle H_i^+ | \phi_2^+ \rangle]$,
and in $\delta g^R$, $\tan\beta$ is replaced by 
$[v_{\rm SM}\langle H_i^+ | \phi_1^+ \rangle]/v_1$.  The charged Higgs
sector can be constrained by $R_b$ when there are no significant contributions
to $Z \rightarrow b \bar{b}$ coming from neutral Higgs 
boson exchange.  This is
ensured when $v_1$ is not too small.  In this regime, $\delta g^L$ can
be significant, while $\delta g^R$ is negligible.
The constraint from $R_b$ on the mass of the remaining light 
charged Higgs boson is the same as in fig.~\ref{fig:mtwohdm}, with
$\tan\beta$ replaced by 
$v_2/[v_{\rm SM}\langle H_i^+ | \phi_2^+ \rangle]$.  

If $v_2$ and $\langle H_i^+ | \phi_2^+ \rangle$ are held constant while
the masses of the heavy charged Higgs bosons are reduced, the bound shown in
fig.~\ref{fig:mtwohdm} becomes stronger.  This happens because the
heavy charged Higgs bosons begin to contribute to $\delta g^L$, forcing the
contribution of the light charged Higgs boson to be smaller in order to be
consistent with the measured value of $R_b$.  This is done by raising the 
mass of the light charged Higgs boson.

Finally, if all the charged Higgs bosons are
degenerate, with a common mass $M_H$, then we can sum over a complete set 
of states and the corrections in a Type II model simplify to the
following:
\begin{eqnarray}
\delta g^L &=& \frac{1}{32 \pi^2} \frac{e}{s_Wc_W} 
\left( \frac{gm_t}{\sqrt{2}M_W} \right)^2 
\frac{v_{\rm SM}^2 - v_2^2}{v_2^2}
 \left[\frac{R}{R-1} - \frac{R \log R}{(R-1)^2}
 \right]\,, 
\end{eqnarray}
\begin{eqnarray}
\delta g^R &=& \frac{1}{32 \pi^2}  \frac{e}{s_Wc_W} 
\left( \frac{gm_b}{\sqrt{2}M_W} \right)^2
\frac{v_{\rm SM}^2 - v_1^2}{v_1^2}
\left[\frac{R}{R-1} - \frac{R \log R}{(R-1)^2}
 \right]\,,
\end{eqnarray}
where $R\equiv m_t^2/M_H^2$.  
These corrections are in addition to the corrections
due to charged Goldstone boson exchange in the SM. 
In a Type I model, $v_2$ is replaced by $v_1$ in
$\delta g^L$.

These corrections are the same as the corrections in the 2HDM, with
$\tan\beta$ replaced by $v_2/(v^2_{\rm SM} - v_2^2)^{1/2}$ in $\delta g^L$,
and $\tan\beta$ replaced by $(v_{\rm SM}^2 - v_1^2)^{1/2}/v_1$ in 
$\delta g^R$.  As before, the charged Higgs
sector can be constrained by $R_b$ when there are no significant contributions
to $Z \rightarrow b \bar{b}$ coming from neutral Higgs boson exchange.  This is
ensured when $v_1$ is not too small.  In this regime,
the constraint from $R_b$ on the common charged Higgs mass $M_H$ 
is the same as in fig.~\ref{fig:mtwohdm}, with 
$\tan\beta$ replaced by $v_2/(v^2_{\rm SM} - v_2^2)^{1/2}$.

\subsubsection{Neutral Higgs boson contributions}
\label{sec:NeutralHiggs}

As we showed in section~\ref{sec-H0}, the radiative corrections to 
the process $Z \rightarrow b \bar{b}$ due to neutral Higgs boson exchange 
are proportional to $\lambda_b^2$.  They
are negligible compared to the contributions from charged Higgs boson
exchange which are proportional to $\lambda_t^2$, except when
$\lambda_b$ is enhanced relative to $\lambda_t$.  This happens
in a Type II model when $v_1\ll v_2$.  In what
follows we consider only Type II models.
%
When $\lambda_b$ is enhanced, the corrections to 
$\delta g^R$ due to charged Higgs boson exchange will also contribute.
These must be taken into account when deriving constraints 
on Higgs sector parameters from the $R_b$ measurement.

\paragraph{Two Higgs doublet model}

The 2HDM contains three neutral Higgs bosons,
\begin{eqnarray}
A^0 &=& -\sin\beta\, \phi_1^{0,i} + \cos\beta\, \phi_2^{0,i}\,,\nonumber  \\
h^0 &=& -\sin\alpha\, \phi_1^{0,r} + \cos\alpha\, \phi_2^{0,r}\,,\nonumber  \\
H^0 &=& \cos\alpha\, \phi_1^{0,r} + \sin\alpha\, \phi_2^{0,r}\,.
\end{eqnarray}
The corrections due to neutral Higgs boson exchange in the 2HDM depend
on the masses of the three neutral Higgs bosons, the mixing
angle $\alpha$, and $\tan\beta \equiv v_2/v_1$.  
The Higgs couplings are easily found from these parameters using
the formulae of section~\ref{sec:extendedHiggs} (see, {\it e.g.},
ref.~\cite{HHG}).  Inserting these couplings into
eq.~(\ref{eq:h0loopsac}) for the corrections from
neutral Higgs boson exchange
\begin{eqnarray}
&&\delta g^{R,L}(a) = \pm \frac{1}{16\pi^2} \frac{e}{s_Wc_W}
  \left( \frac{gm_b}{\sqrt{2}M_W} \right)^2 \tan^2\beta
                        \nonumber \\
&&\qquad \times\left[ \frac{s_{\alpha}}{s_{\beta}}  \cos(\beta - \alpha)
        C_{24}(m_b^2,M_{h^0}^2,M_{A^0}^2)+\frac{c_{\alpha}}{s_{\beta}} 
        \sin(\beta - \alpha) C_{24}(m_b^2,M_{H^0}^2,M_{A^0}^2)
  \right]\,,\nonumber \\
&&\delta g^{R,L}(b) = -\frac{1}{32\pi^2} g^{L,R}_{Zb\bar{b}}
     \left( \frac{gm_b}{\sqrt{2}M_W} \right)^2 \tan^2\beta \nonumber \\
&&\qquad\qquad \times \Biggl[
    \left(\frac{s_{\alpha}}{s_{\beta}}\right)^2
   \Bigl\{\nicefrac{1}{2}-\left[2C_{24} + M_Z^2(C_{22}-C_{23})\right]
   (M_{h^0}^2,m_b^2,m_b^2) \Bigr\}\nonumber \\
&&\qquad\qquad\quad  + \left(\frac{c_{\alpha}}{s_{\beta}}\right)^2
    \Bigl\{\nicefrac{1}{2}-\left[2C_{24} +  M_Z^2(C_{22}-C_{23})\right]
         (M_{H^0}^2,m_b^2,m_b^2)\Bigr\} \nonumber \\[5pt]
&&\qquad\qquad\quad + 
\nicefrac{1}{2}-\left[2C_{24} +  M_Z^2(C_{22}-C_{23})\right]
         (M_{A^0}^2,m_b^2,m_b^2) \Biggr]\nonumber\,, \\[5pt] 
&&\delta g^{R,L}(c) = \frac{1}{32\pi^2} g^{R,L}_{Zb\bar{b}}
  \left( \frac{gm_b}{\sqrt{2}M_W} \right)^2 \tan^2\beta 
      \nonumber \\
&&\times \left[
 \left(\frac{s_{\alpha}}{s_{\beta}}\right)^2
  B_1(m_b^2;m_b^2,M_{h^0}^2)
 + \left(\frac{c_{\alpha}}{s_{\beta}}\right)^2
                        B_1(m_b^2;m_b^2,M_{H^0}^2)
                         + B_1(m_b^2;m_b^2,M_{A^0}^2) \right]
\end{eqnarray}
where $s_{\alpha}\equiv \sin\alpha$, $c_{\alpha}\equiv\cos\alpha$,
$s_{\beta}\equiv\sin\beta$, and $c_{\beta}\equiv\cos\beta$.

The contribution of these corrections to $R_b$ 
can be either positive or negative, depending on the
neutral Higgs masses and the mixing angle $\alpha$.
We plot the corrections for various sets of parameters.


\begin{figure}
\resizebox{\textwidth}{!}{\rotatebox{270}{\includegraphics{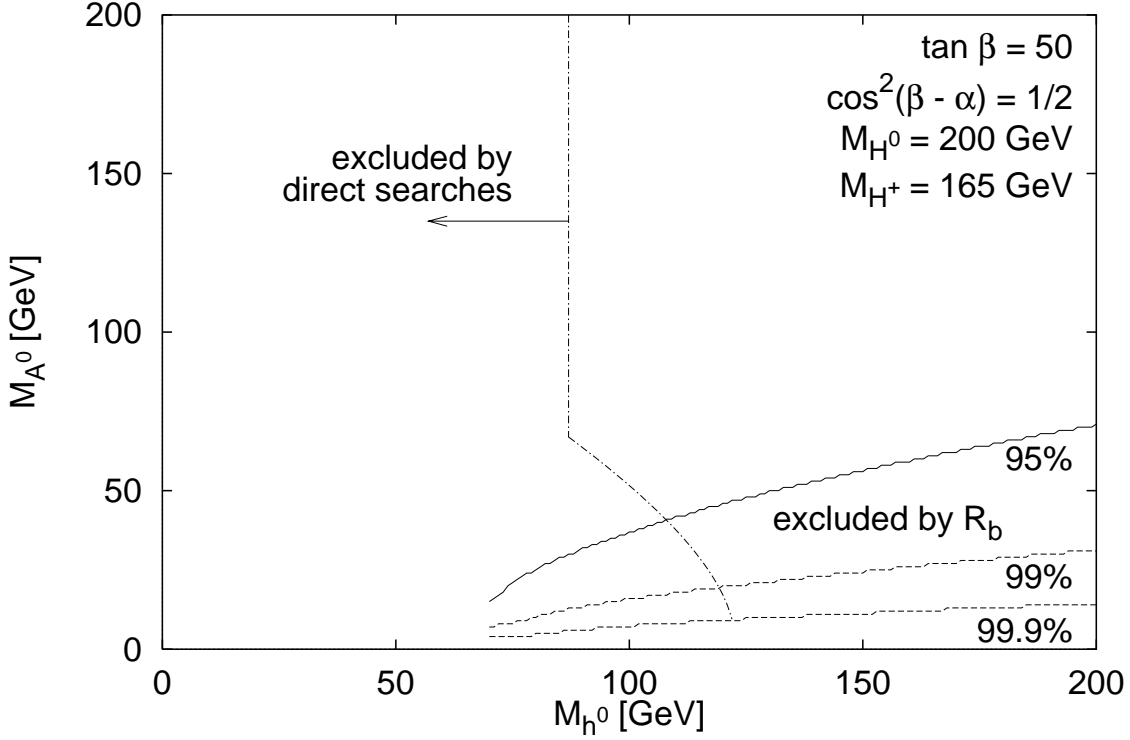}}}
\caption[[$R_b$ in the 2HDM with $\tan\beta = 50$,
$\cos^2(\beta - \alpha) = 1/2$, $M_{H^0} = 200$ GeV, and $M_{H^+} = 165$ GeV]
{$R_b$ in the 2HDM with $\tan\beta = 50$,
$\cos^2(\beta - \alpha) = 1/2$, $M_{H^0} = 200$~GeV and $M_{H^+} = 165$~GeV.
$\Delta R_b < 0$ for all allowed masses, so this model is 
in worse agreement with experiment than the SM.  The solid
line is the 95\% confidence level lower bound on $M_{A^0}$ from
$R_b$.  Also shown are the 99\% and 99.9\% confidence level contours
(dashed lines).
The dot-dashed line is the lower bound on $M_{h^0}$ from direct searches
(see appendix~\ref{app:dirsearches}).
}
\label{fig:Rb0_2_165}
\end{figure}

In fig.~\ref{fig:Rb0_2_165}, we plot 
the constraints on the neutral Higgs sector from $R_b$.  
The parameters in this plot are $\tan\beta = 50$, 
$\cos^2(\beta - \alpha) = 1/2$,
and $M_{H^0} = 200$ GeV.  With $\cos^2(\beta - \alpha) = 1/2$,
the $Zh^0A^0$ and $ZH^0A^0$ couplings are equal, and
$h^0$, $H^0$, and $A^0$ all
contribute to the corrections.  The contribution of the charged Higgs 
boson (which depends on $M_{H^+}$) to $R_b$ must also be considered.  
Note that for large $\tan\beta$, the charged Higgs boson contributions to 
$\delta g^L$ are negligible, whereas contributions to
$\delta g^R$ are negative which reduces $R_b$.
In  fig.~\ref{fig:Rb0_2_165}, we
have taken $M_{H^+}=165$ GeV, which is the lower bound on the
charged Higgs mass in the general 2HDM 
based on constraints from the observed rate for
$b \rightarrow s \gamma$ \cite{CLEO99,Borzumati98}.
We can also consider a second case where $M_{H^+}\gg M_Z$.  In this
limit, the contribution of $H^+$ to $R_b$ vanishes, and we need only
consider the effects of the neutral Higgs sector.\footnote{Since 
$\Delta R_b < 0$ from virtual $H^+$ exchange, in the case of
$M_{H^+} \rightarrow \infty$,
the $R_b$ exclusion contours in fig.~\ref{fig:Rb0_2_165}
would move downward ({\it i.e.}, less parameter space would be
excluded).} 
However, given a
fixed value of $M_{H^0}$, one cannot arbitrarily increase $M_{H^+}$
without violating the constraints due to the $\rho$-parameter.
In appendix~\ref{app:2HDM}, the shift in the $\rho$-parameter due 
to one-loop radiative corrections mediated by
the non-minimal Higgs sector of the 2HDM is given by
eq.~(\ref{rho2HDM}).  As an example, consider the case of
$\cos^2(\beta-\alpha)=1/2$ and $M_{h^0}$, $M_{A^0} \leq M_{H^0}= 200$~GeV 
as in fig.~\ref{fig:Rb0_2_165}.  If we take 
$\Delta\rho\lsim 3\times 10^{-3}$, we find that 
the charged Higgs boson must be lighter than about 270 GeV.  
For $M_{H^+}=270$~GeV, the contour lines in fig.~\ref{fig:Rb0_2_165}
change by an insignificant amount, so there is no need to show a
separate graph.

Since the corrections to $R_b$ from both the charged and neutral Higgs bosons
are proportional to $\tan^2\beta$, we can vary $\tan\beta$ within
the large $\tan\beta$ regime and $\Delta R_b$ will still be negative.
In particular, the region ruled out by $R_b$ gets larger as 
$\tan\beta$ increases.
In fig.~\ref{fig:Rb0_2_165}, the range of masses of $h^0$ and $A^0$ 
in which $\Delta R_b > 0$ is already excluded by direct searches.  
For all remaining allowed $h^0$ and $A^0$ masses,
$\Delta R_b < 0$, in worse agreement with experiment than the SM.
The corresponding
corrections to $A_b$ are negligible ($|\Delta A_b| < 0.003$)
compared to the experimental uncertainty in the $A_b$ measurement.


\begin{figure}
\resizebox{\textwidth}{!}{\rotatebox{270}{\includegraphics{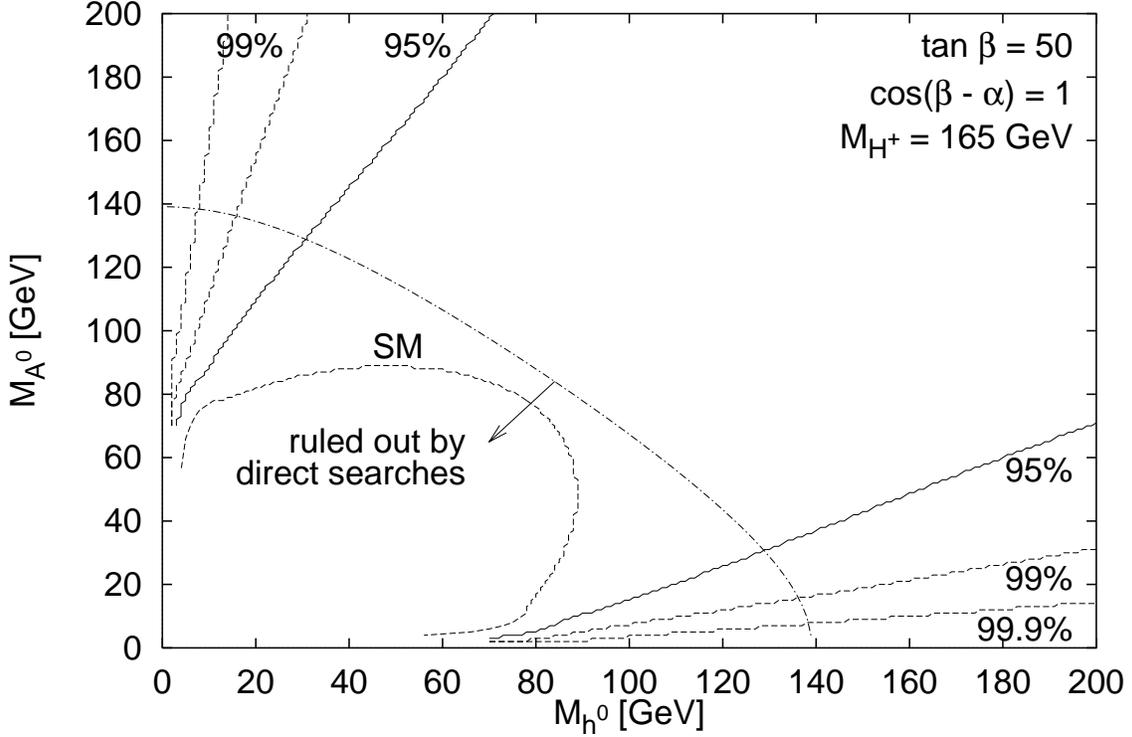}}}
\caption[$R_b$ in the 2HDM with $\tan\beta = 50$, 
$\cos(\beta - \alpha) = 1$, and $M_{H^+} = 165$ GeV]
{$R_b$ in the Type II 2HDM with $\tan\beta = 50$, 
$\cos(\beta - \alpha) = 1$, and $M_{H^+} = 165$~GeV.  The solid lines
are the 95\% confidence level lower bounds on $M_{A^0}$ and $M_{h^0}$
from $R_b$.  The 99\% and 99.9\% confidence level bounds from $R_b$
are also indicated by the appropriately labeled dashed lines.
The (roughly) semi-circular
dashed line labeled ``SM'' is where the predicted value of $R_b$ 
is the same as in the SM.  
The region below this line, in which $\Delta R_b > 0$, is 
entirely excluded by direct searches (see appendix~\ref{app:dirsearches}).
The latter region corresponds to
the area below the dot--dashed line in the direction indicated by the arrow. 
}
\label{fig:Rb0_1_165}
\end{figure}

In fig.~\ref{fig:Rb0_1_165}, we exhibit the 
constraints on the neutral Higgs sector from $R_b$
for $\cos(\beta-\alpha)=1$, with all other parameters the same as
in fig.~\ref{fig:Rb0_2_165}.  For $\cos(\beta - \alpha) = 1$, the
$ZH^0A^0$ coupling is zero and the $H^0 b \bar{b}$ coupling is not 
enhanced over the SM $H^0 b \bar{b}$ coupling, so the contribution of
$H^0$ to the corrections is negligible.  The region where $\Delta R_b > 0$
(due to the positive contribution of the neutral Higgs bosons to
$R_b$ which overcomes the negative contribution from $H^+$ exchange) 
lies below the (roughly) semi-circular dashed contour.  However,
this region of parameter space is already ruled out by the direct
search limits from LEP (see appendix \ref{app:dirsearches}).
Note that the corrections to $R_b$ are negative for large splittings 
between $M_{h^0}$ and $M_{A^0}$.  Thus areas of low $M_{h^0}$ and 
high $M_{A^0}$, and of low $M_{A^0}$ and 
high $M_{h^0}$, are ruled out by the $R_b$ measurement.
Again, the corresponding
corrections to $A_b$ are negligible ($|\Delta A_b| < 0.004$)
compared to the experimental uncertainty in the $A_b$ measurement.  

Both the charged and neutral Higgs boson corrections at large 
$\tan\beta$ are proportional
to $\tan^2\beta$.  Hence, varying $\tan\beta$ will not change the 
combinations of $M_{h^0}$ and $M_{A^0}$ for which $\Delta R_b = 0$.
It follows that the line where $R_b$ is equal to its SM value stays the same
as we vary $\tan\beta$, as long as we remain in the large $\tan\beta$
regime.  Since the corrections grow with $\tan\beta$, the regions ruled out by
$R_b$ in fig.~\ref{fig:Rb0_1_165} get larger as $\tan\beta$ increases.

As previously noted, the corrections from charged Higgs boson exchange
give a negative contribution to $R_b$.  If the charged Higgs mass is 
increased, its negative contribution is reduced, and hence the
excluded regions of fig.~\ref{fig:Rb0_2_165} shrink.  In particular,
the semi-circular contour where $\Delta R_b=0$ moves outward, and
eventually crosses the dot-dashed line (which indicates the boundary of
the region excluded by direct searches).  That is, for large enough
$H^+$ mass, there exists an unexcluded region of the $M_{h^0}$ {\it vs.}
$M_{A^0}$ plane for which $R_b>0$, resulting in a slightly better fit
to the measured value of $R_b$.  However, as noted above, the
charged Higgs mass cannot be taken too large without violating the
$\rho$-parameter constraint.  If this constraint is also imposed,
then even with $M_{H^+}$ taken at its maximally allowed value (with
the parameters as given in fig.~\ref{fig:Rb0_2_165}), the 
viable region of parameter space where $R_b>0$ is quite small.
Moreover, this region is 
on the verge of being ruled out by the direct Higgs searches at LEP.

Finally, we can consider the case of
$\cos(\beta - \alpha)=0$ by interchanging the roles of
$h^0$ and $H^0$ in fig.~\ref{fig:Rb0_1_165};
the results for $R_b$ and $A_b$ will remain the same.  
For $\cos(\beta - \alpha)=0$, the couplings of $h^0$ are equal to 
their SM values, so the SM Higgs search limit applies.
That is, the experimental lower limit of $M_{h^0}$ 
is equivalent to the SM Higgs mass bound from LEP,
$M_{h^0} > 95.2$ GeV \cite{lephiggs}.  
$H^0$ is, by definition, the
heavier CP--even neutral Higgs boson, so $M_{H^0} > M_{h^0} > 95.2$ GeV.
The mass of $H^0$ is also constrained by the LEP search for $H^0 A^0$ 
production. 
When $\cos(\beta - \alpha) = 0$, the $Zh^0A^0$ coupling is zero
and the $h^0b\bar{b}$ coupling is not enhanced over the SM coupling.
Hence, $h^0$ does not contribute significantly to the corrections and we
will neglect it.

\begin{figure}
\resizebox{\textwidth}{!}{\rotatebox{270}{\includegraphics{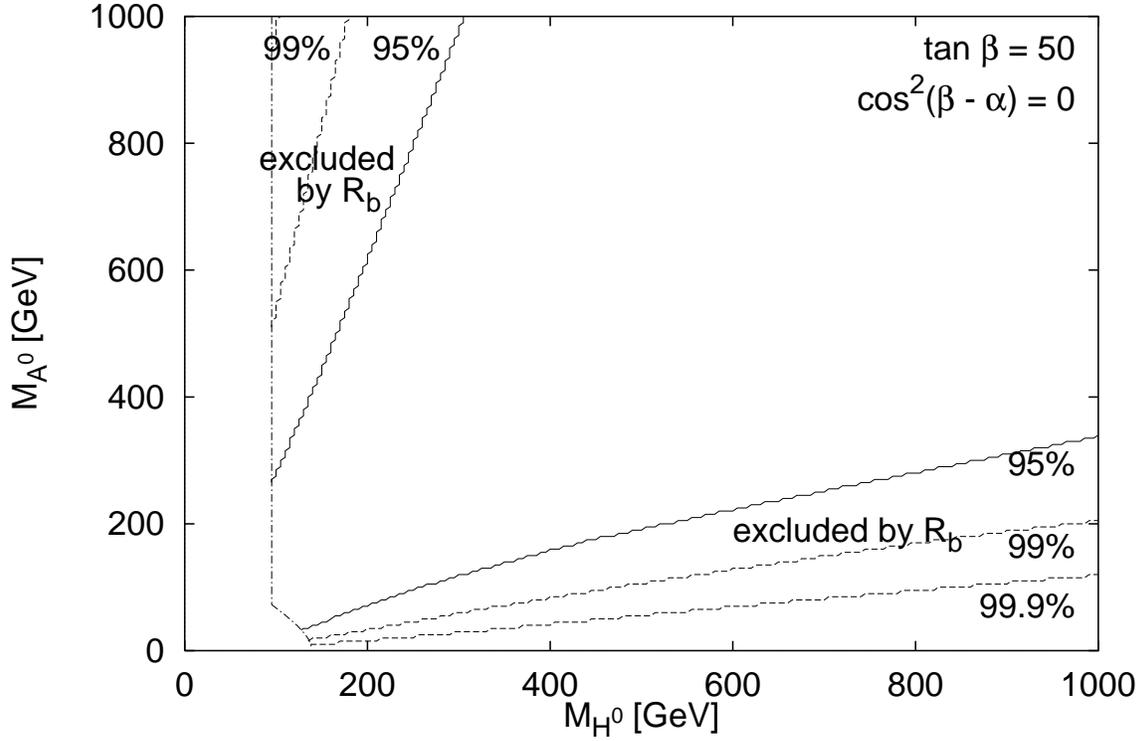}}}
\caption[$R_b$ in the 2HDM with $\tan\beta = 50$ and
$\cos(\beta - \alpha) = 0$]
{$R_b$ in the 2HDM with $\tan\beta = 50$ and 
$\cos(\beta - \alpha) = 0$.
For $M_{A^0}>165$ GeV, we take $M_{H^+}=M_{A^0}$,
while for $M_{A^0} < 165$ GeV, we take $M_{H^+}=165$~GeV.
The solid lines are the 95\% confidence level lower bounds on $M_{A^0}$
and $M_{H^0}$ from $R_b$.  The labeled dashed lines are the
99\% and 99.9\% confidence level bounds from $R_b$.
The dot--dashed line is the bound from direct searches
(see appendix \ref{app:dirsearches}).
}
\label{fig:Rb0_0_no}
\end{figure}

The constraints on the Higgs parameters from $R_b$ for 
$\cos(\beta-\alpha)=0$ are shown in fig.~\ref{fig:Rb0_0_no}.
To ensure that the $\rho$-parameter is satisfied, we 
have set $M_{H^+} = M_{A^0}$ for $M_{A^0} > 165$ GeV.  For
$M_{A^0} < 165$ GeV, we have taken $M_{H^+} = 165$ GeV, 
to be consistent with the constraint from
$b \rightarrow s \gamma$ \cite{CLEO99,Borzumati98}.
The $R_b$ measurement rules out areas of parameter space
where the mass splitting between $H^0$ and $A^0$ is large.  
For example, if the $H^0$ ($A^0$) mass is 1000 GeV, then $A^0$ ($H^0$) 
must be heavier than about 300 GeV.

\paragraph{Two Higgs doublet model in the decoupling limit}

In the decoupling limit of the 2HDM, $h^0$ remains light
and its couplings to the SM particles approach those of the SM Higgs boson, 
while all the other Higgs bosons become heavy and nearly degenerate in mass. 
In particular, in the decoupling limit \cite{decoupling}: 
(i) $M_{h^0} \sim \mathcal{O}(M_Z)$; (ii) $M_{H^0} \simeq M_{A^0}\simeq
M_{H^+}\gg M_Z$; (iii) $| M^2_{H^0} - M^2_{A^0} |  \sim
| M^2_{H^+} - M^2_{A^0} | \sim \mathcal{O}(M_Z^2)$; and
(iv) $\cos (\beta - \alpha) \sim \mathcal{O}(M_Z^2/M_{A^0}^2)$.
We can expand the corrections to $Z \rightarrow b \bar{b}$ from neutral
Higgs boson exchange in the 2HDM in this limit.  Expanding 
the three--point integrals in the limit of $M_{A^0}\gg M_Z$
(see, {\it e.g.}, ref.~\cite{PHI}), we obtain
to leading order in $M_Z^2/M^2_{A^0}$:
\begin{eqnarray}
&&\delta g^L  \simeq  \frac{1}{16 \pi^2} \left( \frac{e}{s_Wc_W} \right) 
  \left( \frac{g m_b}{\sqrt{2} M_W} \right)^2 \tan^2 \beta\, 
\frac{M_Z^2}{M_{A^0}^2}   
    \left\{ -\frac{1}{36} + \frac{1}{9} s^2_W 
    \left[\frac{1}{3} + \log \left(-\frac{M^2_{A^0}}{M_Z^2} \right) 
                                   \right]\right\}\,,\nonumber   \\
&&\delta g^R \simeq \frac{1}{16 \pi^2} \left( \frac{e}{s_Wc_W} \right) 
    \left( \frac{g m_b}{\sqrt{2} M_W} \right)^2 \tan^2 \beta \nonumber \\
 && \qquad \quad  \times \frac{M_Z^2}{M_{A^0}^2}  
      \left\{ -\frac{1}{36} - \frac{1}{6}\left[ \log
      \left(-\frac{M^2_{A^0}}{M_Z^2} \right)\right]
     + \frac{1}{9} s^2_W  \left[ \frac{1}{3} 
     + \log \left(-\frac{M^2_{A^0}}{M_Z^2} \right)\right] \right\}\,.
\end{eqnarray}
As an example, for $\tan\beta = 50$ and $M_{A^0} = 200$ GeV, 
the above corrections give $\Delta R_b = -3.7\times 10^{-4}$, 
which is only half the size of the experimental 
error of the $R_b$ measurement.  The corrections vanish in the limit
of large $M_{A^0}$ as expected from decoupling.  This limit
is approached in fig.~\ref{fig:Rb0_0_no} 
when $M_{H^0}$ and $M_{A^0}$ are both large (compared to $M_Z$) 
and similar in size.

\paragraph{Multiple--doublet models}

We now consider neutral Higgs boson exchange in a model containing
multiple Higgs doublets, denoted $\Phi_k$, with hypercharge $Y=1$.

In a Type I model of this type, we let $\Phi_1$ couple to both up--
and down--type quarks, and none of the other doublets couple to quarks.
In a Type II model, we let $\Phi_1$ couple only to down--type quarks, 
and $\Phi_2$ couple to up--type quarks.  Then the Yukawa couplings 
are defined in the same way as in the 2HDM, in 
eqs.~(\ref{eqn:lambdaI})--(\ref{eqn:lambdaII}).  As always, the 
contributions to $Z \rightarrow b \bar{b}$ from neutral Higgs boson exchange
are only significant in a Type II model, when $\lambda_b$ is enhanced
by small $v_1$.  We will only consider Type II multi--doublet models
with small $v_1$.

The contributions from neutral Higgs boson exchange in the multi--doublet
model are more complicated than in the 2HDM, simply because there are
more neutral Higgs states.  Only the states which have a nonzero overlap with
$\Phi_1$ can couple to $b$ quarks, so only these states contribute.
The corrections depend on the overlap of each neutral state with $\Phi_1$
and the mass of each state.
As in the 2HDM, the region of parameter space in which the 
correction to $R_b$ is positive is almost entirely ruled out by 
direct searches.

\paragraph{Multiple--doublet models with Higgs singlets}

We can also consider adding a number of Higgs singlets, with hypercharge
zero, to the multi--doublet model.  The singlets do not couple to 
$Z$ or to quarks.  Their vevs are also unconstrained by the $W$ mass.
In general, the singlets will mix with the neutral components of the 
doublets to form mass eigenstates.  
The couplings
of the physical states to $b\bar{b}$ still depend only on $v_1$, which
fixes $\lambda_b$, and on the overlap of each state with $\Phi_1$.
The couplings of physical states to $Z$ are no longer the same as in
a model containing only doublets.  Instead, they are equal to the 
$Z$ coupling for doublet states weighted by the overlap of each state
with doublets.  Explicitly,
\begin{equation}
g_{ZH^0_iA^0_j} = \frac{-ie}{2s_Wc_W} \sum_{k}
                        \langle H_i^0 | \phi_k^{0,r} \rangle
                        \langle A_j^0 | \phi_k^{0,i} \rangle,
\label{eqn:ZHAwithsinglets}
\end{equation}
where $k$ runs only over the Higgs doublets.

In order to understand the effects of singlets on the corrections
to $Z \rightarrow b \bar{b}$, let us imagine replacing each Higgs 
singlet with the neutral component of a doublet, with the appropriate 
CP quantum number, while holding the masses and mixings of the 
physical states fixed.  Then, the couplings of each state to 
$b \bar{b}$ remain the same.  However, the couplings of the states
to $Z$ are now equal to,
\begin{equation}
g_{ZH^0_iA^0_j} = \frac{-ie}{2s_Wc_W},
\end{equation}
which is the coupling in a model containing only Higgs doublets.
Comparing this to eq.~(\ref{eqn:ZHAwithsinglets}), we see that
$\delta g^{R,L}(a)$ in the model with singlets must be smaller in 
magnitude than in the model in which the singlets are replaced by 
doublets.

\paragraph{Degenerate neutral Higgs bosons in a general extended
Higgs sector}

The corrections to $Z \rightarrow b \bar{b}$ due to neutral Higgs boson
exchange in a general model are quite complicated.  
They depend on the couplings and masses of all the 
neutral Higgs bosons in the model.  However, the corrections can 
be simplified if some of the neutral Higgs bosons are degenerate in 
mass.  

Consider a general extended Higgs sector, 
which can contain Higgs singlets, doublets, and larger multiplets.
We require that the model be Type II, and that $\lambda_b$ 
be enhanced relative to $\lambda_t$.  A Type II model must
contain at least two Higgs doublets, $\Phi_1$ and $\Phi_2$, 
to couple to down-- and up--type quarks, respectively.
Only the neutral Higgs bosons with large couplings to $b\bar{b}$ give
significant contributions to the corrections.  In what follows 
we will only consider these.  States without enhanced $b\bar{b}$ 
couplings, such as $G^0$, do not contribute significantly.  We
will ignore them, and therefore it does not matter what their
masses are.

If all the CP--even neutral Higgs bosons are degenerate with mass $M_H$, 
and all the CP--odd neutral Higgs bosons are degenerate with mass $M_A$,
then we can take the two-- and three--point integrals outside of
the sums in eq.~(\ref{eq:h0loopsac}).
Then we can sum the couplings over complete sets of states.  
Using the couplings in a general model from
eqs.~(\ref{eq:gV}), (\ref{eq:gA}) and (\ref{eq:ZHA}), we find
\begin{eqnarray}
&&\sum_{H^0_i,A^0_j} g_{ZH^{0}_{i}A^{0}_{j}} g^{V}_{H^{0}_{i}b\bar{b}} 
                        g^{A}_{A^{0}_{j}b\bar{b}}
                        = \frac{e}{2s_Wc_W} 
                        \left( \frac{gm_b}{\sqrt{2}M_W} \right)^2 
                        \left( \frac{v_{\rm SM}}{v_1} \right)^2\,,
                        \\
&&\sum_{H^0_i} (g^{V}_{H^{0}_{i}b\bar{b}})^{2}
                        = -\sum_{A^0_j} (g^{A}_{A^{0}_{j}b\bar{b}})^{2}
                        = \left( \frac{gm_b}{\sqrt{2}M_W} \right)^2
                        \left( \frac{v_{\rm SM}}{v_1} \right)^2.
\end{eqnarray}
These sums over the couplings are related to certain couplings in
the 2HDM, as follows.  
On the left--hand side are the couplings in the general model with degenerate 
neutral Higgs bosons, and on the right--hand side are the couplings 
in the 2HDM with $\cos(\beta - \alpha) = 1$.  That is,
\begin{eqnarray}
\sum_{H^0_i,A^0_j} g_{ZH^{0}_{i}A^{0}_{j}} g^{V}_{H^{0}_{i}b\bar{b}} 
                        g^{A}_{A^{0}_{j}b\bar{b}}
                        &=& g_{Zh^0A^0} g^{V}_{h^0b\bar{b}} 
                        g^{A}_{A^0b\bar{b}}\,,
                        \\
\sum_{H^0_i} (g^{V}_{H^{0}_{i}b\bar{b}})^{2}
                        &=& (g^{V}_{h^0b\bar{b}})^{2}\,,
                        \\
\sum_{A^0_j} (g^{A}_{A^{0}_{j}b\bar{b}})^{2}
                        &=& (g^{A}_{A^0b\bar{b}})^{2}\,.
\end{eqnarray}

Therefore, when all the CP--even neutral Higgs bosons are degenerate 
with mass $M_H$, and all the CP--odd neutral Higgs bosons are degenerate 
with mass $M_A$, the contributions to $Z \rightarrow b\bar{b}$ are
the same as the contributions from the 2HDM with $M_{h^0} = M_H$,
$M_{A^0} = M_A$, and $\cos(\beta - \alpha) = 1$.  The parameter
corresponding to $\tan\beta$ in the extended model is
\begin{equation}
 \frac{v^2_{\rm SM} - v^2_1}{v^2_1}  = \tan^2\beta.
\end{equation}
                        %

Similarly, the corrections can be simplified if only the CP--even
states, or only the CP--odd states, are degenerate.  If the 
CP--even states are degenerate, we can sum over the $H^0_i$ 
couplings.  We then get the same
result as if the CP--even neutral Higgs sector consisted 
of a single state $H^0$, which consists entirely of $\phi_1^{0,r}$.
Recall that $\phi_1^{0,r}$ is the CP--even neutral component of
the doublet which couples to down--type quarks.
If, instead, the CP--odd states are degenerate, we can sum over the
$A^0_j$ couplings.  We get the same result as if the CP--odd
neutral Higgs sector consisted of a single state $A^0$, which
consists entirely of $\phi_1^{0,i}$ (up to the small mixing of
$\phi_1^{0,i}$ with $G^0$, which is negligible in the small $v_1$
regime).

\subsection{Models with Higgs multiplets larger than doublets}
\label{sec:exotics}

We next consider Higgs sectors that contain one or more multiplets 
larger than doublets.
Two types of models that use different approaches
to satisfy $\rho \simeq 1$ are examined.  
We first consider models in which the
vevs of the multiplets larger than doublets are fine-tuned to be very
small, so that their contribution to the $\rho$ parameter is 
negligible.  Second, we consider models that preserve ${\rm SU}(2)_c$
symmetry (in the Higgs sector), ensuring that $\rho = 1$ at tree level.

%

\subsubsection{Models with one Higgs doublet and one triplet}
\label{sec:finetuned}

The minimal extension 
of the Higgs sector that includes multiplets larger than doublets
consists of the complex $Y=1$ doublet of the SM,
denoted by $\Phi$, plus a triplet Higgs field.  The vev of the triplet
field must be fine--tuned very small in order to be consistent
with the measured value of the $\rho$ parameter, $\rho \simeq 1$.
The triplet field can either be a real triplet with
$Y=0$, or a complex triplet with $Y=2$.  Here we investigate both 
possibilities.

These two models contain only one Higgs doublet, which couples to 
both up-- and down--type quarks, so they are necessarily
Type I models.  Thus $\lambda_b \ll \lambda_t$, and the only 
non--negligible contributions to $Z \rightarrow b \bar{b}$ come
from the contributions to $\delta g^L$ from charged Higgs boson exchange.


We first consider the ``$Y=0$ model'' with one doublet and one real triplet 
field with $Y=0$.  
The triplet field is
$\xi = (\xi^{+}, \xi^{0}, \xi^{-})$.  We define the doublet and 
triplet vevs by $\langle \phi^0 \rangle = v_{\phi}/\sqrt{2}$ and
$\langle \xi^0 \rangle = v_{\xi}$.  The vevs are constrained by the
$W$ mass to satisfy
\begin{equation}
v_{\rm SM}^2 = v_{\phi}^2 + 4 v_{\xi}^2\,.
\end{equation}
It is convenient to parameterize
the ratio of the vevs by
\begin{equation}
\tan \theta_0 = \frac{v_{\phi}}{2 v_{\xi}}\,.
\end{equation}
In this model, the tree--level $\rho$ parameter is
\begin{equation}
\rho = \frac{v_{\phi}^{2} + 4 v_{\xi}^{2}}{v_{\phi}^{2}}
     = 1 + \frac{4 v_{\xi}^{2}}{v_{\phi}^{2}} \equiv 1 + \Delta \rho\,.
\end{equation}
In terms of $\tan\theta_0$, we find
\begin{equation}
\Delta \rho = \frac{1}{\tan^2 \theta_0}.
\end{equation}
We see that in order to have $\rho \simeq 1$, the triplet vev
must be very small, giving large $\tan\theta_0$.
The charged states mix to form the charged Goldstone boson and a
single charged physical state
\begin{eqnarray}
G^{+} &=& \sin\theta_0\, \phi^{+} + \cos\theta_0\, \xi^{+}\nonumber\,,  \\
H^{+} &=& \cos\theta_0\, \phi^{+} - \sin\theta_0\, \xi^{+}\,.
\end{eqnarray}


We next consider the ``$Y=2$ model'' with one doublet and one complex triplet
field with $Y=2$.  
The triplet field is
$\chi = ( \chi^{++}, \chi^+, \chi^0 )$.  We define the vev of this 
triplet field by $\langle \chi^0 \rangle = v_{\chi}/\sqrt{2}$.
The vevs are constrained by the
$W$ mass to satisfy
\begin{equation}
v_{\rm SM}^2 = v_{\phi}^2 + 2 v_{\chi}^2\,.
\end{equation}
It is convenient to parameterize the ratio of the doublet and triplet
vevs by
\begin{equation}
\tan \theta_2 = \frac{v_{\phi}}{\sqrt{2} v_{\chi}}\,.
\end{equation}
In this model, the tree--level $\rho$ parameter is
\begin{equation}
\rho = \frac{ v_{\phi}^2 + 2 v_{\chi}^2 }{ v_{\phi}^2 + 4 v_{\chi}^2 }
    = 1 - \frac{ 2 v_{\chi}^2 }{ v_{\phi}^2 + 4 v_{\chi}^2 }
                        \equiv 1 + \Delta \rho\,.
\end{equation}
In terms of $\tan\theta_2$, we find
\begin{equation}
\Delta \rho = \frac{-1}{\tan^2 \theta_2 + 2}.
\end{equation}
We see that in order to have $\rho \simeq 1$, the triplet vev
must be very small, giving large $\tan\theta_2$.
The charged states mix to form the charged Goldstone boson and a
single charged physical state
\begin{eqnarray}
G^{+} &=& \sin\theta_2\,\phi^{+} + \cos\theta_2\, \xi^{+}\,,\nonumber  \\
H^{+} &=& \cos\theta_2\,\phi^{+} - \sin\theta_2\, \xi^{+}\,.
\end{eqnarray}

The Higgs couplings to quarks and the $Z$ boson can be parameterized
as follows.  We let $\theta$ denote $\theta_0$ in the $Y=0$ model and
$\theta_2$ in the $Y=2$ model.  We also define a factor $\epsilon$
such that $\epsilon = +1$ in the $Y=0$ model and $\epsilon = -1$ 
in the $Y=2$ model.  The charged Higgs couplings to quarks are
\begin{eqnarray}
g^{L}_{G^{+}\bar{t}b} &=& \frac{gm_t}{\sqrt{2}M_W}\,,  \\
g^{L}_{H^{+}\bar{t}b} &=& \frac{gm_t}{\sqrt{2}M_W} \cot\theta\,.
\end{eqnarray}
The $ZH^+_iH^-_j$ couplings are
\begin{eqnarray}
g_{ZG^{+}G^{-}} &=& -\frac{e}{s_{W}c_{W}} \left(\frac{1}{2} - s^{2}_{W}
                        + \frac{\epsilon}{2}\cos^2 \theta \right)\,,  \\
g_{ZG^{+}H^{-}} &=& \frac{e}{s_{W}c_{W}} \frac{\epsilon}{2} \sin\theta 
                        \cos\theta \,, \\
g_{ZH^{+}H^{-}} &=& -\frac{e}{s_{W}c_{W}} \left(\frac{1}{2} - s^{2}_{W}
                        + \frac{\epsilon}{2} \sin^2 \theta \right)\,.
\end{eqnarray}
                        %

\paragraph{Contributions to \boldmath $Z \rightarrow b\bar{b}$}

In both the $Y=0$ and the $Y=2$ models, there is an off--diagonal
$ZG^{+}H^{-}$ coupling, and the diagonal $ZH^{+}H^{-}$ and 
$ZG^{+}G^{-}$ couplings differ from their values in models containing
only Higgs doublets and singlets.  These couplings contribute
to the second and third terms of $\delta g^L$ in 
eq.~(\ref{eq:h+loops2}).

In addition to the SM contribution to $\delta g^L_{\rm SM}$ from $G^+$ 
exchange, the charged Higgs contribution to $\delta g^L$ is given by
\begin{eqnarray}
&&\delta g^L = \frac{1}{32 \pi^2} \left( \frac{g m_t}{\sqrt{2} M_W} \right)^2
 \frac{e}{s_Wc_W} \cos^2 \theta 
 \biggl\{ \frac{1}{\sin^2\theta} 
 \left[\frac{R}{R-1} - \frac{R \log R}{(R-1)^2} \right]
                         \nonumber \\[6pt]
 &&\!\!\!\!\!\!  -  2 \epsilon \left[ 
                         C_{24} (m_t^2, M_W^2, M_W^2) 
                  + C_{24} (m_t^2, M_{H^+}^2, M_{H^+}^2)
                 - 2 C_{24} (m_t^2, M_W^2, M_{H^+}^2) \right] \biggr\}
\label{eqn:dgL1doub1trip}
\end{eqnarray}
where $R \equiv m_{t}^{2}/M_{H^+}^2$.  

Note that $\delta g^L$ is proportional to $\cos^2\theta$, which
goes to zero in the large $\tan\theta$ limit.  This is due to the
fact that in the limit of small triplet vev in either of these models, 
the overlap of $H^+$ with the doublet is proportional to $\cos\theta$.
As a result, in the large $\tan\theta$ limit,
$H^+$ is almost entirely triplet and so its couplings to 
quarks are very small.  Also in the large $\tan\theta$ limit,
the off--diagonal $ZG^+H^-$ coupling goes to zero, and the 
$ZG^+G^-$ coupling approaches its SM value.

\paragraph{Constraints from the $\rho$ parameter}

We must also take into account the constraint on $\tan\theta$ 
from the $\rho$ parameter in each of the models.  Since 
$\Delta \rho$ depends differently on $\tan\theta_0$ than on
$\tan\theta_2$, the constraint on $\tan\theta$ will be different
in the $Y=0$ model than in the $Y=2$ model.

The experimental constraints on $\Delta \rho$ are taken from 
ref.~\cite{Erler99}.  Writing the tree-level value of the
$\rho$-parameter as $\rho=1+\Delta\rho_{\rm new}$, the 2$\sigma$ level
limits are:
\begin{equation} \label{rhobounds}
-1.7 \times 10^{-3} < \Delta \rho_{\rm new} < 2.7 \times 10^{-3}\,.
\end{equation}
We now use $\Delta \rho_{\rm new}$ to constrain $\tan\theta_0$ and 
$\tan\theta_2$.  We ignore the radiative corrections from the non--minimal
Higgs sector.
In the $Y=0$ model, $\Delta \rho_{\rm new}>0$, while in the
$Y=2$ model, $\Delta \rho_{\rm new}<0$.
The resulting $2\sigma$ limits on $\tan\theta_0$ and 
$\tan\theta_2$ are:
\begin{equation}
\tan\theta_0 > 19\,,\qquad  \tan\theta_2 > 24\,.
\end{equation}

\paragraph{Results}

The contribution to $\delta g^L$ in both the $Y=0$ model and
the $Y=2$ model is proportional to $\cos^2\theta$ 
[eq.~(\ref{eqn:dgL1doub1trip})].  When the constraints on $\tan\theta$
from the $\rho$ parameter are imposed, the corrections to $R_b$ and $A_b$
are very small.  Even allowing for the largest possible values of
$\theta_0$ and $\theta_2$, we find that over the relevant Higgs
parameter space (with $M_{H^+}$ varying between 10 and 1000 GeV), 
$|\Delta R_b|< 7\times 10^{-6}$ and  $|\Delta A_b|< 3\times 10^{-6}$.
These corrections are tiny compared to the experimental error 
on the $R_b$ and $A_b$ measurements 
[eqs.~(\ref{eqn:Rbmeasured}) and (\ref{eqn:Abmeasured})].

In general,
the contribution to $\delta g^L$ vanishes in the large $\tan\theta$
limit in any model in which the charged
Goldstone boson is made up almost entirely of the doublet
that couples to quarks.
As a result, the overlap of the other charged
Higgs states with the doublet is very small, so the other charged
Higgs states couple very weakly to quarks.  This 
occurs in any model that contains only one scalar doublet,
plus any number of singlets and multiplets larger than doublets,
as long as the vevs of the multiplets larger than doublets are
forced to be small.

The contributions of multiplets larger than doublets to 
$Z \rightarrow b\bar{b}$ can be large only if the larger 
multiplets mix significantly with doublets, so that the 
resulting Higgs states have non--negligible couplings to quarks.
This can happen in two ways.  First, if the model contains more than 
one doublet, then each singly-charged scalar field 
of the doublets that couples to quarks will contain physical scalar
components that can mix with charged scalar states from higher multiplets.
The resulting physical charged scalar mass eigenstates
can thus possess a non--negligible couplings to quarks.  A model 
of this type is discussed in section~\ref{sec:2doub1trip}.
Second, if the multiplets larger than doublets have sizeable
vevs, then the charged Goldstone boson must contain some admixture
of the larger multiplets, leaving part of the doublet free to 
mix into the physical charged Higgs states.  However, in order for the
multiplets larger than doublets to have sizeable vevs without
violating the constraint from the $\rho$ parameter, the model 
must preserve ${\rm SU}(2)_c$ symmetry.  Models of this type are 
discussed in section~\ref{sec:triplets}.

\subsubsection{Models with two doublets and one triplet}
\label{sec:2doub1trip}

We next consider a Higgs sector consisting of two doublets and one triplet.  
As in section~\ref{sec:finetuned},
the triplet can be real with $Y=0$ or complex with $Y=2$.
The couplings for these models are given in ref.~\cite{hlthesis}.
With two doublets, we can construct either a Type I model or a Type II model.
In this section we consider a Type II model, but we also note the changes in
the formulae that must be made to recover a Type I model.

We will consider both the corrections due to charged Higgs boson exchange and
the corrections due to neutral Higgs boson exchange.  
The corrections from neutral
Higgs boson exchange can be significant in a Type II model with large
$\tan\beta$.
We define $\tan\beta$ in this model exactly as in the 2HDM,
$\tan\beta = v_2/v_1$, where the vevs of the doublets are 
$\langle \phi_1^0 \rangle = v_1/\sqrt{2}$ and 
$\langle \phi_2^0 \rangle = v_2/\sqrt{2}$.

\paragraph{Charged Higgs boson contributions}

We first consider the corrections due to charged Higgs boson exchange
in the ``$Y=0$ model'' consisting of two doublets and one real triplet 
field with $Y=0$.  The triplet field is
$\xi = (\xi^{+}, \xi^{0}, \xi^{-})$.  We define the 
triplet vev by
$\langle \xi^0 \rangle = v_{\xi}$.
In the $Y=0$ model we parameterize the vevs by
\begin{equation}
\tan \theta_0 = \frac{(v_1^2 + v_2^2)^{1/2}}{2 v_\xi}\,,
\end{equation}
in analogy to section~\ref{sec:finetuned}.

The charged Higgs states are defined as follows.
The Goldstone boson is
\begin{equation}
G^+ = \sin\theta_0 (\cos\beta\, \phi_1^+ + \sin\beta\, \phi_2^+) 
                        + \cos\theta_0\, \xi^+\,.
\end{equation}
In addition we define two orthogonal states
\begin{eqnarray}
H_1^{+\prime} &=& \cos\theta_0 (\cos\beta\, \phi_1^+ + \sin\beta\, \phi_2^+) 
                        - \sin\theta_0\, \xi^+\,,\nonumber \\
H_2^{+\prime} &=& -\sin\beta\, \phi_1^+ + \cos\beta\, \phi_2^+\,,
\end{eqnarray}
which will mix by an angle $\delta$ to form the mass eigenstates.  
Before mixing
them, however, let us take the limit of large $\tan\theta_0$ 
in order to satisfy the experimental constraint on the $\rho$-parameter.  
We make the approximation that $\sin\theta_0 \approx 1$ and
$\cos\theta_0 \approx 0$ (the general case of arbitrary 
$\tan\theta_0$ is considered in 
ref.~\cite{hlthesis}).  Then the positively charged scalar states are
\begin{eqnarray}
G^+ &\simeq& \cos\beta\, \phi_1^+ + \sin\beta\, \phi_2^+\,,\nonumber \\
H_1^{+\prime} &\simeq& -\xi^+\,,\nonumber \\
H_2^{+\prime} &=& -\sin\beta\, \phi_1^+ + \cos\beta\, \phi_2^+\,,.
\end{eqnarray}
These states mix by an angle $\delta$ to form the mass eigenstates:
\begin{eqnarray}
H_1^+ &\simeq& \sin\delta\, (-\sin\beta\, \phi_1^+ + \cos\beta\, \phi_2^+)
                        - \cos\delta\, \xi^+\,,\nonumber \\
H_2^+ &\simeq& \cos\delta\, (-\sin\beta\, \phi_1^+ + \cos\beta\, \phi_2^+)
                        + \sin\delta\, \xi^+\,.
\end{eqnarray}

We next consider the corrections due to charged Higgs boson exchange
in the ``$Y=2$ model'' consisting of two doublets and one complex triplet field
with $Y=2$.  The triplet field is
$\chi = ( \chi^{++}, \chi^+, \chi^0 )$.  We define the triplet
vev by $\langle \chi^0 \rangle = v_{\chi}/\sqrt{2}$.
In the $Y=2$ model we parameterize the vevs by
\begin{equation}
\tan\theta_2 = \frac{(v_1^2 + v_2^2)^{1/2}}{\sqrt{2} v_{\chi}}\,,
\end{equation}
again in analogy to section~\ref{sec:finetuned}.

The charged Higgs states in the $Y=2$ model are parameterized in
the same way as the states in the $Y=0$ model.
The Goldstone boson is
\begin{equation}
G^+ = \sin\theta_2\, (\cos\beta\, \phi_1^+ + \sin\beta\, \phi_2^+) 
                        + \cos\theta_2\, \chi^+\,.
\end{equation}
In addition we define two orthogonal states:
\begin{eqnarray}
H_1^{+\prime} &=& \cos\theta_2\, (\cos\beta\, \phi_1^+ + \sin\beta\, \phi_2^+) 
                        - \sin\theta_2\, \chi^+\,,\nonumber \\
H_2^{+\prime} &=& -\sin\beta\, \phi_1^+ + \cos\beta\, \phi_2^+\,,
\end{eqnarray}
which will mix by an angle $\delta$ to form the mass eigenstates.  
Before mixing
them, we shall take the limit of large $\tan\theta_2$ 
in order to satisfy
the experimental constraint on the $\rho$-parameter.  We make the 
approximation $\sin\theta_2 \approx 1$ and $\cos\theta_2 \approx 0$
(the general case of arbitrary $\tan\theta_2$ is considered in 
ref.~\cite{hlthesis}).  Then the positively charged scalar states are
\begin{eqnarray}
G^+ &\simeq& \cos\beta\, \phi_1^+ + \sin\beta\, \phi_2^+\,,\nonumber \\
H_1^{+\prime} &\simeq& - \chi^+\,,\nonumber \\
H_2^{+\prime} &=& -\sin\beta\, \phi_1^+ + \cos\beta\, \phi_2^+\,.
\end{eqnarray}
These states mix by an angle $\delta$ to form the mass eigenstates:

\begin{eqnarray}
H_1^+ &\simeq& \sin\delta\, (-\sin\beta\, \phi_1^+ + \cos\beta\, \phi_2^+)
                        - \cos\delta\, \chi^+\,,\nonumber \\
H_2^+ &\simeq& \cos\delta\, (-\sin\beta\, \phi_1^+ + \cos\beta\, \phi_2^+)
                        + \sin\delta \chi^+\,.
\end{eqnarray}
The states and couplings for arbitrary $\tan\theta_2$ are given
in ref.~\cite{hlthesis}.

We now calculate the
corrections to $Z \rightarrow b\bar{b}$ from charged
Higgs boson exchange in the Type II $Y=0$ and $Y=2$ models.
As in section~\ref{sec:finetuned}, we introduce the parameter 
$\epsilon = +1$ in 
the $Y=0$ model, and $\epsilon = -1$ in the $Y=2$ model.
In addition to the SM correction due to charged Goldstone boson exchange,
the charged Higgs contributions to $\delta g^L$ are given by
                        %
\begin{eqnarray}
&&\delta g^L  \simeq 
 \frac{1}{32 \pi^2} \frac{e}{s_Wc_W}
       \left( \frac{g m_t}{\sqrt{2} M_W} \right)^2 
        \cot^2 \beta \nonumber \\
&&\qquad \times \left\{\sin^2 \delta 
      \left[ \frac{R_1}{R_1-1} 
      -\frac{R_1 \log R_1}{(R_1-1)^2} \right]
      + \cos^2 \delta
      \left[ \frac{R_2}{R_2-1} 
      - \frac{R_2 \log R_2}{(R_2-1)^2} \right] \right\}
                                  \nonumber \\
  &&\qquad  - \frac{\epsilon}{16 \pi^2} \frac{e}{s_Wc_W}
  \left( \frac{g m_t}{\sqrt{2} M_W} \right)^2 \cot^2 \beta 
  \sin^2 \delta \cos^2 \delta \nonumber \\  
&&\!\!\! \times  \left[ C_{24}(m_t^2,M_{H_1^+}^2,M_{H_1^+}^2) 
         + C_{24}(m_t^2,M_{H_2^+}^2,M_{H_2^+}^2) 
          - 2 C_{24}(m_t^2,M_{H_1^+}^2,M_{H_2^+}^2)\right]
\label{eqn:dgL2doub1trip}
\end{eqnarray}
where $R_i\equiv m_t^2 / M_{H_i^+}^2$.  In the Type I models,
$\delta g^L$ is the same as above with $\cot^2\beta$ replaced by
$\tan^2\beta$.

The first term of eq.~(\ref{eqn:dgL2doub1trip})
is the same as the correction in a three Higgs doublet
model (3HDM), given in eq.~(\ref{eqn:dgLMHDM}).
It is 
positive, which gives a negative contribution to $R_b$, taking
it farther from the measured value.  The second
term comes from the effects of the triplet.  This second term is 
proportional to $\sin^2\delta \cos^2\delta$, so it is only 
significant for $\delta$ near $\pi/4$, which corresponds to maximal
mixing between the charged doublet and triplet states in $H_1^+$ and
$H_2^+$.  The second term is zero if $H_1^+$ and $H_2^+$ have the 
same mass.  

The sign of the second term depends on the hypercharge
of the Higgs triplet.  In the $Y=0$ model,
the second term is negative.
However, the second term is smaller in magnitude than the first
term, so the overall contribution to $\delta g^L$ is positive in 
the $Y=0$ model.
In fig.~\ref{fig:D2T_0+}, we plot the constraints on 
$M_{H_1^+}$ and $M_{H_2^+}$ from the $R_b$ measurement in the 
$Y=0$ model, for
maximal doublet--triplet mixing ($\delta = \pi/4$) and $\tan\beta = 1$.
In order for the $Y=0$ model with maximal doublet--triplet mixing
to be consistent with the $R_b$ measurement,
one or both of the charged Higgs bosons must be very heavy.  
\begin{figure}
\resizebox{\textwidth}{!}{\rotatebox{270}{\includegraphics{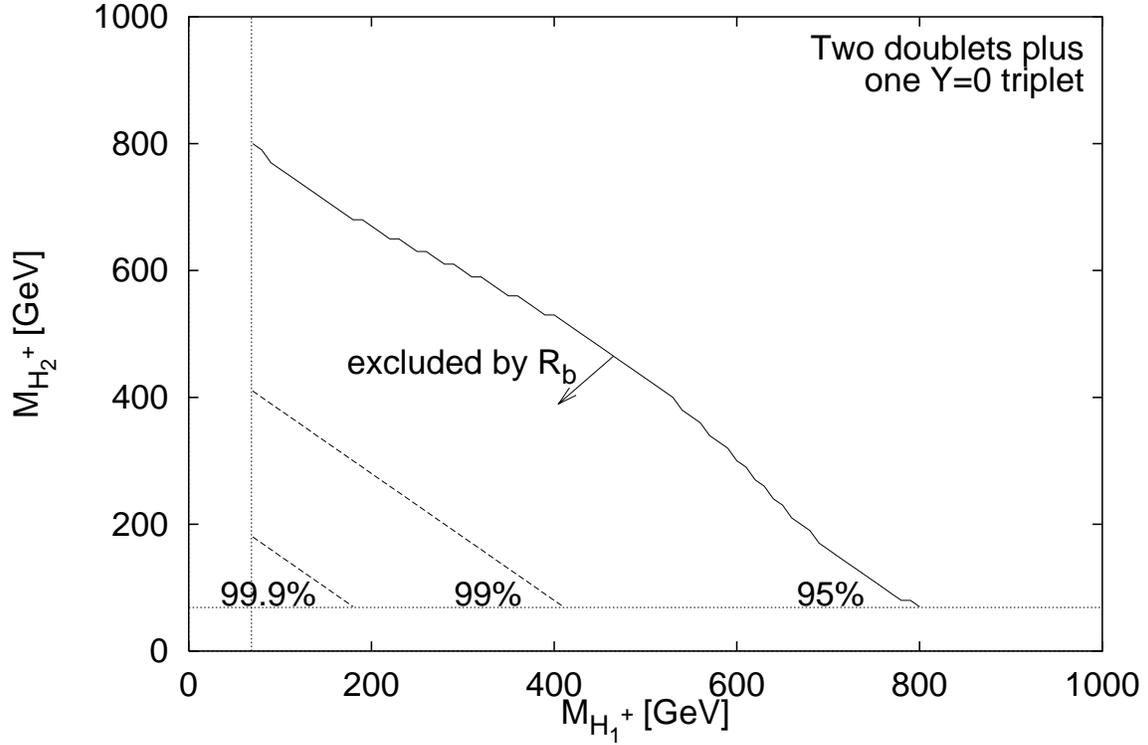}}} 
\caption[Constraints from $R_b$ on the masses of the two charged Higgs
states in the model with two doublets and one real
$Y=0$ triplet]
{Constraints from $R_b$ on the masses of the two charged Higgs
states $H_1^+$ and $H_2^+$ in the model with two doublets and one real
$Y=0$ triplet, with $\tan\beta = 1$ and $\delta = \pi/4$.  The
area below the solid line is excluded at 95\% confidence level.  Also
shown are the 99\% and 99.9\% confidence levels (dashed).  
The dotted lines correspond to the LEP lower limit for the $H^+$ mass,
$M_{H^+} > 77.3$~GeV \cite{lephiggs}.
} 
\label{fig:D2T_0+} 
\end{figure}

In the $Y=2$ model, the second term of eq.~(\ref{eqn:dgL2doub1trip})
is positive, resulting in a 
positive $\delta g^L$ which is larger than in the $Y=0$ model.
As a result, a larger area of parameter space is excluded by the
$R_b$ measurement in the
$Y=2$ model than in the $Y=0$ model.
In fig.~\ref{fig:D2T_2+}, we
plot the constraints on $M_{H_1^+}$ and $M_{H_2^+}$ from the
$R_b$ measurement on the $Y=2$ model, for maximal doublet--triplet
mixing ($\delta = \pi/4$) and $\tan\beta = 1$.  From the $R_b$ 
constraint with these parameters, we find that both of the 
charged Higgs bosons must be 
heavier than 410 GeV.  If $\delta$ is varied or $\tan\beta$ is 
increased, this bound becomes lower.  Note that we do not plot
a direct search bound on the $H^+$ mass.  In this model, the LEP bound
on the charged Higgs mass does not apply, as explained 
in appendix~\ref{app:dirsearches}.
\begin{figure}
\resizebox{\textwidth}{!}{\rotatebox{270}{\includegraphics{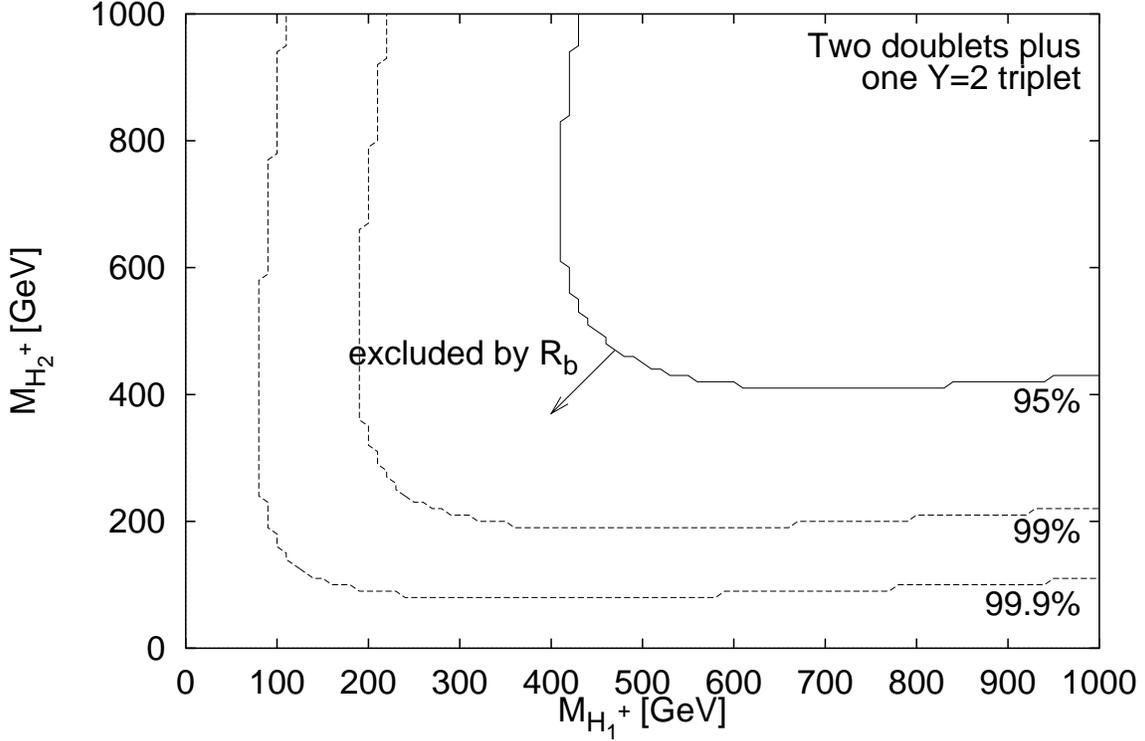}}}
\caption[Constraints from $R_b$ on the masses of the two charged Higgs
states in the model with two doublets and one complex
$Y=2$ triplet]
{Constraints from $R_b$ on the masses of the two charged Higgs
states $H_1^+$ and $H_2^+$ in the model with two doublets and one complex
$Y=2$ triplet, with $\tan\beta = 1$ and $\delta = \pi/4$.  The
area below the solid line is excluded at 95\% confidence level.  For these
values of $\tan\beta$ and $\delta$, $H^+$ masses below 410 GeV are ruled
out. Also
shown are the 99\% and 99.9\% confidence levels (dashed).
}
\label{fig:D2T_2+}
\end{figure}

For completeness, we also write the contributions to $\delta g^R$, which 
are only significant at large $\tan\beta$. 
For both the Type I and Type II models 
                        %
\begin{eqnarray}
&&\delta g^R \simeq - \frac{1}{32 \pi^2}  \frac{e}{s_Wc_W}
        \left( \frac{g m_b}{\sqrt{2} M_W} \right)^2 
        \tan^2 \beta \nonumber \\
&&\qquad \times \left\{ \sin^2 \delta 
    \left[ \frac{R_1}{R_1-1} - \frac{R_1 \log R_1}{(R_1-1)^2} \right]
    + \cos^2 \delta \left[ \frac{R_2}{R_2-1} 
     - \frac{R_2 \log R_2}{(R_2-1)^2} \right] \right\} \nonumber \\
&&\qquad  - \frac{\epsilon}{16 \pi^2} \frac{e}{s_Wc_W}
  \left( \frac{g m_b}{\sqrt{2} M_W} \right)^2 \tan^2 \beta 
   \sin^2 \delta \cos^2 \delta   \nonumber \\
&&\!\!\times \left[ C_{24}(m_t^2,M_{H_1^+}^2,M_{H_1^+}^2) 
    + C_{24}(m_t^2,M_{H_2^+}^2,M_{H_2^+}^2) 
   - 2 C_{24}(m_t^2,M_{H_1^+}^2,M_{H_2^+}^2) \right]
\label{eqn:dgR2doub1trip}
\end{eqnarray}
where $\epsilon = +1$ in the $Y=0$ model and $\epsilon = -1$ 
in the $Y=2$ model.
The first term of eq.(\ref{eqn:dgR2doub1trip})
is the same as the correction in a 3HDM.  The second 
term comes from the effects of the triplet.

\paragraph{Neutral Higgs boson contributions}

Consider the contributions to $Z \rightarrow b \bar{b}$ 
from neutral Higgs boson exchange in these models.  The corrections can
only be significant in the Type II models when $\tan\beta$ is large.
For this reason, we disregard the Type I models here.

In the $Y=0$ model, there is no $Z \xi^0 A^0$ coupling \cite{hlthesis}.
As a result, $\xi^0$ has the  same couplings as a Higgs singlet.
Thus, the corrections from neutral Higgs boson exchange 
have the same form as in a model containing two doublets
and a real singlet with $Y=0$.  Models of this type were discussed 
in section~\ref{sec:NeutralHiggs}. 

In the $Y=2$ model, there are nonzero $Z \chi^{0,r} \chi^{0,i}$ 
couplings \cite{hlthesis}.
The neutral Higgs states can be written in the 
large $\tan\theta_2$ limit as
\begin{eqnarray}
H_1^0 &=& \cos\gamma\, (\cos\alpha\, \phi_1^{0,r} 
        + \sin\alpha\, \phi_2^{0,r}) + \sin\gamma\, \chi^{0,r}\,,
                        \label{eqn:2d1tH10}\\
H_2^0 &=& -\sin\gamma\, (\cos\alpha\, \phi_1^{0,r} 
        + \sin\alpha\, \phi_2^{0,r}) + \cos\gamma\, \chi^{0,r}\,,  
                        \label{eqn:2d1tH20} \\
H_3^0 &=& -\sin\alpha\, \phi_1^{0,r} + \cos\alpha\, \phi_2^{0,r}\,,
                        \label{eqn:2d1tH30}   \\
G^0 &\simeq& \cos\beta\, \phi_1^{0,i} + \sin\beta\, \phi_2^{0,i}\,,  \\
A_1^0 &\simeq& - \sin\omega \sin\beta\, \phi_1^{0,i} 
                         + \sin\omega \cos\beta\,
                        \phi_2^{0,i} - \cos\omega\, \chi^{0,i}\,,  \\
A_2^0 &\simeq& - \cos\omega \sin\beta\,
                        \phi_1^{0,i}
                        + \cos\omega \cos\beta\,
                        \phi_2^{0,i} + \sin\omega \chi^{0,i}\,,
\end{eqnarray}
where, for simplicity, only $H_1^0$ and $H_2^0$ contain triplet
admixtures.  We find that
the contributions of the neutral Higgs bosons in this model 
can be split into two pieces.  The first piece is the same as the
contribution in a 3HDM, in which
the neutral Higgs states are given as above but with
the triplet states $\chi^{0,r}$ and $\chi^{0,i}$ replaced by the
neutral states of the third doublet.  This piece is denoted
by $\delta g^{R,L}_{\rm 3HDM}$.  The second piece contains
the additional contribution due to the effects of the isospin and 
hypercharge
of the triplet, and is denoted $\delta g^{R,L}_{\rm triplet}$.
That is,
\begin{equation}
\delta g^{R,L} = \delta g^{R,L}_{\rm 3HDM} + \delta g^{R,L}_{\rm triplet}\,.
\end{equation}

Explicit formulae can be found in ref.~\cite{hlthesis}.  One finds that
$\delta g^{R,L}_{\rm triplet}$ is only significant near maximal 
doublet--triplet mixing in both the CP--odd and CP--even sectors,
which occurs when $\omega$ and $\gamma$ are both near $\pm \pi/4$.
In addition, $\delta g^{R,L}_{\rm triplet}$ is zero if 
$M_{H_1^0} = M_{H_2^0}$ or $M_{A_1^0} = M_{A_2^0}$.  Its sign depends 
on the mixing angles and the Higgs masses.  For all the neutral Higgs
bosons lighter than about 200~GeV and maximal doublet--triplet 
mixing, the contribution to $R_b$ from $\delta g^{R,L}_{\rm triplet}$ is 
smaller than the contribution to $R_b$ from 
$\delta g^{R,L}_{\rm 3HDM}$ over most of the parameter space.
The contribution to $R_b$ from $\delta g^{R,L}_{\rm 3HDM}$ is of
the same order of magnitude as the contribution to $R_b$ from the 
neutral sector of the 2HDM.


\subsubsection{Georgi--Machacek model with \boldmath${\bf SU}(2)_c$ symmetry}
\label{sec:triplets}

In order to obtain $\rho = 1$ at tree level the electroweak symmetry
breaking must preserve a ``custodial'' SU(2) symmetry, called 
${\rm SU}(2)_c$, that ensures equal masses are given to the $W^{\pm}$ and
$W^{3}$.  We refer to models with this property as generalized
Georgi--Machacek (G--M) models, after the extended model of this type
with Higgs triplets first introduced in ref.~\cite{Georgi1}.

The triplet G--M model contains a complex
$Y=1$ doublet $\Phi$, a real $Y=0$ triplet $\xi$, and a complex 
$Y=2$ triplet $\chi$.
The Higgs fields take the form
\begin{equation}
\Phi = \left( \begin{array}{cc}
                                  \phi^{0*} & \phi^{+} \\
                                  -\phi^{+*}  & \phi^{0}
                              \end{array}   \right)
\end{equation}
\begin{equation}
\chi = \left(  \begin{array}{ccc}
                                  \chi^{0*}  & \xi^{+} & \chi^{++} \\
                                  -\chi^{+*}  & \xi^{0} & \chi^{+}  \\
                                  \chi^{++*} & \xi^{-} & \chi^{0}
                               \end{array}  \right)
\end{equation}
where $\xi^- = -(\xi^+)^*$,
which transform under ${\rm SU}(2)_{L} \times {\rm SU}(2)_{R}$ 
as $(\half,\half)$ 
and $(1,1)$ representations, respectively.  
The electroweak symmetry breaking preserves
${\rm SU}(2)_{c}$ when the vevs of the fields are 
diagonal, $\langle \chi \rangle = v_{\chi} \mathbf{I}$ and 
$\langle \phi^0 \rangle = (v_{\phi} / \sqrt{2}) \mathbf{I}$, where
$\mathbf{I}$ is the unit matrix.
The vevs are constrained by the $W$ mass to satisfy
\begin{equation}
v_{\rm SM}^2=v_\phi^2+8v_\chi^2\,.
\end{equation}
It is convenient to parameterize the ratio of vevs by
\begin{equation}
\tan\theta_H \equiv \frac{2\sqrt{2} v_{\chi}}{v_{\phi}}\,.
\end{equation}

Under the electroweak symmetry breaking, the 
${\rm SU}(2)_{L} \times {\rm SU}(2)_{R}$
symmetry is broken down to ${\rm SU}(2)_c$.  A representation $(T,T)$ of 
${\rm SU}(2)_{L} \times {\rm SU}(2)_{R}$ 
decomposes into a set of representations
of ${\rm SU}(2)_c$, in particular,
$2T \oplus 2T-1 \oplus \cdots \oplus 1 \oplus 0$.  In the triplet
G--M model, $\Phi$ breaks down to a 
triplet and a singlet of ${\rm SU}(2)_c$, and $\chi$ breaks down to 
a five-plet, a triplet, and a singlet of ${\rm SU}(2)_c$.
The $W^{\pm}$ and $Z$ bosons are given mass by absorbing the 
${\rm SU}(2)_{c}$
triplet of Goldstone bosons, $G_{3}^{+,0,-}$.
The remaining physical states are
a five-plet $H_{5}^{++,+,0,-,--}$, a triplet
$H_{3}^{+,0,-}$, and two singlets $H_{1}^{0}$ and $H_{1}^{0\prime}$.
If the Higgs potential is chosen to preserve ${\rm SU}(2)_{c}$, then states 
transforming in different representations of ${\rm SU}(2)_{c}$
cannot mix, and the states in each representation are degenerate.
This model contains only one doublet $\Phi$ which gives mass to both the
top- and bottom-type quarks.  Therefore it is a Type I model and
$\lambda_b \ll \lambda_t$.  
Thus the only sizeable correction to the $Z b \bar{b}$ vertex in this 
model will come from the left-handed charged Higgs boson loops.

The two singly-charged Higgs bosons and $G^{+}$ can be written in terms
of the combinations of triplet fields
\begin{equation}
\psi^{+} = \frac{1}{\sqrt{2}} (\chi^{+} - \xi^{+})\,,
\end{equation}
which transforms in a triplet of ${\rm SU}(2)_{c}$, and
\begin{equation}
\zeta^{+} = \frac{1}{\sqrt{2}} (\chi^{+} + \xi^{+})\,,
\end{equation}
which transforms in a five-plet of ${\rm SU}(2)_{c}$.
Then, the singly charged Higgs bosons are
\begin{equation}
G^{+}_{3} = c_H \phi^{+} + s_H \psi^{+}\,,
\end{equation}
\begin{equation}
H^{+}_{3} = -s_H \phi^{+} + c_H \psi^{+}\,,
\end{equation}
\begin{equation}
H^{+}_{5} = \zeta^{+}\,,
\end{equation}
where $s_H\equiv\sin\theta_H$ and $c_H\equiv\cos\theta_H$.

If the Higgs potential is chosen to preserve ${\rm SU}(2)_c$ then
$H^+_3$ and $H^+_5$ are mass eigenstates because they transform
under different representations of ${\rm SU}(2)_c$ \cite{Chanowitz1}. 
Such a potential is desirable because it preserves
${\rm SU}(2)_{c}$ (and $\rho=1$) to all orders in the Higgs
self--couplings.  However, renormalization
of the parameters in the Higgs potential at one loop  
introduces quadratically divergent terms that break ${\rm SU}(2)_{c}$
\cite{Gunion2}.  These terms lead to quadratically divergent contributions
to the $\rho$-parameter and to the mixing of some of the Higgs states,
including $H^{+}_{3}$ and $H^{+}_{5}$.  In order to cancel the divergent
corrections, ${\rm SU}(2)_{c}$--breaking 
counterterms must be introduced in the 
bare Lagrangian and fine--tuned to restore $\rho \simeq 1$.  These 
${\rm SU}(2)_{c}$--violating 
corrections arise at the two--loop level in $R_{b}$,
so they will be neglected here.

The couplings in this model have been given in 
refs.~\cite{Gunion1,HHG}.  
They are also derived in ref.~\cite{hlthesis} for a general
G--M model containing
one multiplet $\Phi = (\half,\half)$ and one larger multiplet $X = (T,T)$.
The doublet field $\Phi$ is the only field with
quark Yukawa couplings.  Under ${\rm SU}(2)_c$ the doublet decomposes into a
singlet and a triplet.  Thus only ${\rm SU}(2)_c$ singlets and triplets can 
contain a doublet admixture and couple to quarks.  This is a general 
feature of any model whose Higgs sector obeys a custodial ${\rm SU}(2)_{c}$
symmetry.  
In the triplet G--M model the charged Higgs couplings to 
quarks are
\begin{equation}
g^{L}_{G^{+}\bar{t}b} = \frac{gm_{t}}{\sqrt{2}M_{W}}\,,
\end{equation}
\begin{equation}
g^{L}_{H_{3}^{+}\bar{t}b} = \frac{-gm_{t}}{\sqrt{2}M_{W}} \tan\theta_H\,,
\end{equation}
\begin{equation}
g^{L}_{H_{5}^{+}\bar{t}b} = 0\,.
\end{equation}
These couplings also hold in a general G--M model containing
$\Phi = (\half,\half)$ and $X = (T,T)$, if $\tan\theta_H$ is defined as 
\begin{equation}
\tan\theta_H = \frac{v_X \sqrt{\frac{4}{3}T(T+1)(2T+1)}}{v_{\phi}}\,,
\end{equation}
where the vevs are constrained by the W mass to satisfy
\begin{equation}
v_{\rm SM}^2=v_\phi^2+\nicefrac{4}{3}T(T+1)(2T+1)v_X^2\,.
\end{equation}
The loop corrections to $R_{b}$ will only involve the charged Higgs
states that appear in the 
triplet representations of ${\rm SU}(2)_{c}$; namely,
$H^{+}_{3}$ and $G^{+}$.

The relevant $ZH^{+}H^{-}$ couplings for charged Higgs bosons in a 
triplet of ${\rm SU}(2)_c$ for any model which preserves 
${\rm SU}(2)_c$ are given below:
\begin{eqnarray}
g_{ZG^{+}G^{-}} &=& \frac{-e}{s_{W}c_{W}}   \left(\half -
                        s^{2}_{W}\right)\,,\nonumber \\[6pt]
g_{ZG^{+}H_{3}^{-}} &=& 0\,, \nonumber \\[6pt]
g_{ZH_{3}^{+}H_{3}^{-}} &=& \frac{-e}{s_{W}c_{W}} 
                        \left(\half - s^{2}_{W}\right)\,,
\end{eqnarray}
as shown in ref.~\cite{hlthesis}.
The loop corrections to $R_{b}$ involving $H^{+}$ are particularly
simple because the $ZG^{+}H_{3}^{-}$ coupling is zero.

In any model which preserves ${\rm SU}(2)_c$ and contains only two multiplets
$\Phi$ and $X$, the correction to $\delta g^L$ is (not including
the SM correction due to the charged Goldstone loops):
\begin{equation}
\delta g^{L} = \frac{1}{32\pi^{2}} 
      \left(\frac{gm_{t}}{\sqrt{2}M_{W}} \right)^2 \tan^{2}\theta_H
      \frac{e}{s_{W}c_{W}} 
     \left[ \frac{R}{R-1} - \frac{R \log R}{(R-1)^{2}} \right]\,,
\end{equation}
from loops involving $H_{3}^{+}$, where $R \equiv 
m_{t}^{2}/M_{H_{3}^{+}}^{2}$.  This correction is positive definite and 
has the same form as the correction in the 2HDM (equation
\ref{eq:dgl2HDM}).  

In general for a model with custodial ${\rm SU}(2)_{c}$ 
and more than one exotic
multiplet $X$, the correction becomes
\begin{equation}
\delta g^{L} =  \frac{1}{32\pi^{2}} \sum_{H_{3i}^{+}} 
     (g^{L}_{H_{3i}^{+} \bar{t}b})^{2} \frac{e}{s_{W}c_{W}}
     \left[\frac{R_{i}}{R_{i}-1} - \frac{R_{i} \log R_{i}}{(R_{i}-1)^{2}}
             \right]\,,
\end{equation}
which is positive definite.
Thus when the Higgs potential is invariant under ${\rm SU}(2)_{c}$, 
the corrections always decrease $R_{b}$.

As in the 2HDM, the $R_{b}$ measurement can be used to set a lower bound
on the mass of the ${\rm SU}(2)_{c}$ triplet $H_{3}$, which varies with 
$\tan \theta_H$.  This bound is independent of the isospin of the 
exotic ${\rm SU}(2)_L \times {\rm SU}(2)_R$ 
multiplet $X$ (or $\chi$ in the triplet
G--M model).
In fig.~\ref{fig:tripcustRb} we plot the bound on $M_{H_3}$ as a 
function of $\tan\theta_H$.

For $H_3$ lighter than about 1 TeV, the $R_b$ measurement
implies that $\tan\theta_H < 2$.
In the triplet G--M model, this corresponds to an
upper limit on the triplet vev of $v_{\chi}/v_{\phi} < 0.7$.
As in the Type I 2HDM, the charged Higgs boson contribution to 
$b \rightarrow s \gamma$ is small compared to the 
contribution in the Type II 2HDM \cite{Grinstein90}, and the 
$b \rightarrow s \gamma$ measurement does not provide additional bounds 
on the parameter space. 
In contrast, for the parameter regions considered above,   
the correction to $A_b$ is negligible ($|\Delta A_b| < 0.002$)
compared to the experimental uncertainty in the $A_b$ measurement.  

\begin{figure}
\resizebox{\textwidth}{!}{\rotatebox{270}{\includegraphics{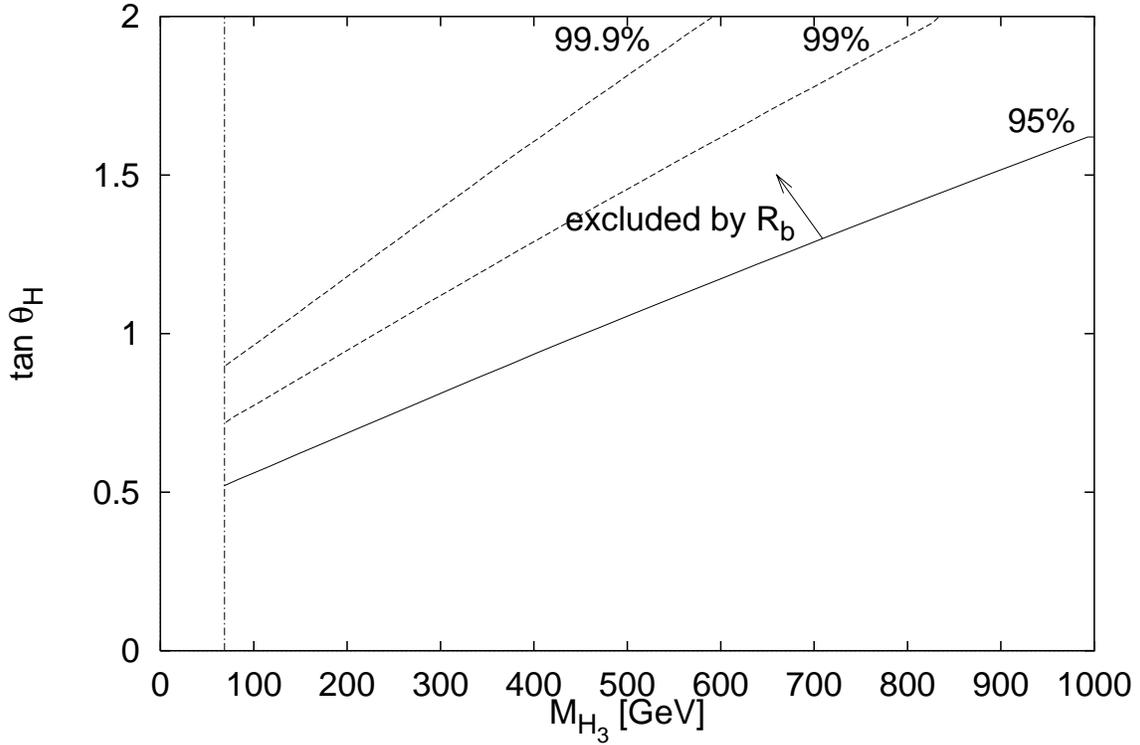}}}
\caption[Bounds from $R_b$ for the G--M model with 
Higgs triplets]
{Bounds from $R_b$ for the G--M model with Higgs triplets
and ${\rm SU}(2)_c$ symmetry.  
The area above the solid line is ruled out at 95\% confidence
level by $R_b$.  Also shown (top to bottom) are the 99.9\% and 99\%
confidence level contours (dashed).  
The dot-dashed line corresponds to the LEP lower limit for the $H^+$ mass,
$M_{H^+} > 77.3$~GeV \cite{lephiggs}.
}
\label{fig:tripcustRb}
\end{figure}

\paragraph{Higgs potential without \boldmath${\bf SU}(2)_c$ invariance}

If the requirement of ${\rm SU}(2)_c$ symmetry is relaxed, it is no longer
meaningful to write the Higgs fields with 
${\rm SU}(2)_L \times {\rm SU}(2)_R$ matrices.
In the triplet model we must define the vevs of the two 
${\rm SU}(2)_L$ triplets separately,
$\langle \chi^0 \rangle = v_{\chi}$, and $\langle \xi^0 \rangle = v_{\xi}$.
Then ${\rm SU}(2)_c$ symmetry corresponds to $v_{\chi} = v_{\xi}$.
The triplet model
can still satisfy $\rho = 1$ if the Higgs potential is fine-tuned so 
that $v_{\chi} = v_{\xi}$.  In this situation the two physical charged
Higgs bosons $H_{3}^{+}$ and $H_{5}^{+}$ can mix with each other.  If we 
parameterize this mixing with an angle $\alpha$, the new mass eigenstates
are
\begin{eqnarray}
H_{1}^{+} &=& \sin\alpha\, H_{3}^{+} + \cos\alpha\,
H_{5}^{+}\,,\nonumber \\
H_{2}^{+} &=& \cos\alpha\, H_{3}^{+} - \sin\alpha\, H_{5}^{+}\,.
\end{eqnarray}
The charged Higgs couplings to the $Z$ and quark pairs are
\begin{eqnarray}
g^{L}_{H_{1}^{+}\bar{t}b}
& =& \frac{gm_{t}}{\sqrt{2}M_{W}} \tan\theta_H \sin\alpha \,,\nonumber \\
g^{L}_{H_{2}^{+}\bar{t}b} 
&=&= \frac{gm_{t}}{\sqrt{2}M_{W}} \tan\theta_H \cos\alpha\,,\nonumber \\
g_{ZG^{+}H_{1}^{-}} &=& \frac{-e}{s_{W}c_{W}} \frac{1}{2} s_H \cos\alpha
\,,\nonumber \\
g_{ZG^{+}H_{2}^{-}} &=& \frac{e}{s_{W}c_{W}} \frac{1}{2} s_H \sin\alpha
\,,\nonumber \\
g_{ZH_{1}^{+}H_{1}^{-}} &=& \frac{-e}{s_{W}c_{W}} (\frac{1}{2} - s^{2}_{W}
     - c_H \sin\alpha \cos\alpha)\,,\nonumber \\
g_{ZH_{1}^{+}H_{2}^{-}} &=& \frac{-e}{s_{W}c_{W}} \frac{1}{2} 
        c_H (\sin^{2}\alpha - \cos^{2}\alpha)\,,\nonumber \\
g_{ZH_{2}^{+}H_{2}^{-}} &=& \frac{-e}{s_{W}c_{W}} (\frac{1}{2} - s^{2}_{W}
                                  + c_H \sin\alpha \cos\alpha)\,.
\end{eqnarray}

Both of the singly charged Higgs bosons 
couple to quarks instead of just one.
There are now off--diagonal $ZH^+_iH^-_j$ couplings with $i\neq j$ 
and non--SM--like terms 
in the diagonal couplings which contribute to $\delta g^L$.  We find:
\begin{eqnarray}
&&\delta g^{L}_{H^{+}} = \delta g^{L}_{G^{+}} (\rm SM)  \nonumber \\
   &&\qquad + \frac{1}{32 \pi^{2}} \frac{e}{s_{W}c_{W}}
\left( \frac{gm_{t}}{\sqrt{2}M_{W}} \right)^{2} \tan^2\theta_H
   \nonumber \\
&&\qquad\qquad \times \left\{ \sin^{2}\alpha \left[ \frac{R_{1}}{R_{1}-1}
   - \frac{R_{1} \log R_{1}}{(R_{1}-1)^{2}} \right]
  + \cos^{2}\alpha \left[ \frac{R_{2}}{R_{2}-1}
   - \frac{R_{2} \log R_{2}}{(R_{2}-1)^{2}} \right]
     \right\}           \nonumber \\
&&\qquad + \frac{1}{16 \pi^{2}} \left(\frac{e}{s_{W}c_{W}} \right)
   \left( \frac{gm_{t}}{\sqrt{2}M_{W}}   \right)^{2} \tan^2\theta_H 
    (2 c_H \sin\alpha \cos\alpha)   \nonumber \\
 &&\qquad\qquad \times \left\{ C_{24}(m_{t}^2,M_{W}^2,M_{2}^2) 
      - C_{24}(m_{t}^2,M_{W}^2,M_{1}^2)   \right.   \nonumber \\[5pt]
 &&\qquad\qquad + \sin^{2}\alpha [C_{24}(m_{t}^2,M_{1}^2,M_{1}^2)
       - C_{24}(m_{t}^2,M_{1}^2,M_{2}^2)]       \nonumber \\[5pt]
 &&\qquad\qquad + \left. \cos^{2}\alpha [C_{24}(m_{t}^2,M_{1}^2,M_{2}^2)
     - C_{24}(m_{t}^2,M_{2}^2,M_{2}^2)] \right\},
\label{eq:233mixed}
\end{eqnarray}
where $R_{i}\equiv m_{t}^{2}/M_{i}^{2}$.
The first term is the SM correction due to $G^{+}$.  The second term is 
positive definite and has the same mass dependence as the charged Higgs boson
correction in the 2HDM.  The third term arises from the off-diagonal
$ZH^{+}H^{-}$ couplings and the non--SM parts of the diagonal $ZH^{+}H^{-}$ 
couplings.  This third term can be positive or negative, depending on the
mixing angle $\alpha$.  It is negative for $M_{H_{2}^{+}} > M_{H_{1}^{+}}$ 
when $\sin\alpha \cos\alpha$ is positive, and grows with increasing 
splitting between $M_{H_{1}^{+}}$ and $M_{H_{2}^{+}}$ and between 
$M_{W}$ and the charged Higgs masses.  

This model is fine-tuned to $v_{\chi} = v_{\xi}$ to give $\rho =1$; 
when the parameters of the Higgs potential are renormalized
this fine-tuning will be lost.  In order to satisfy the 
experimental bounds on $\Delta \rho_{\rm new}$ [eq.~(\ref{rhobounds})],
we must have 
\begin{equation}
- 1.7 \times 10^{-3} < \Delta \rho_{\rm new} 
        = \frac{4(v_{\xi}^2 - v_{\chi}^2)}{v_{\phi}^2 + 8 v_{\chi}^2}
       < 2.7 \times 10^{-3}
\end{equation}
or $ -(5.1~\mathrm{GeV})^2 < v_{\xi}^2 - v_{\chi}^2 < (6.4~\mathrm{GeV})^2 $.
For the model to
be ``natural'' we require the parameters to be of the same order as their 
fine-tuning, or $v_{\chi} \sim v_{\xi} \sim 6$ GeV. Then 
the correction to the SM result in eq.~(\ref{eq:233mixed}) is 
suppressed by a factor of $\tan^2\theta_H \sim 0.005$.

\section{Conclusions}
\label{sec:conclusions}

Radiative corrections to the process $Z \rightarrow b\bar{b}$
arise in extended Higgs sectors due to the exchange of the
additional singly--charged and neutral Higgs bosons.
Because the radiative corrections affect the predictions for $R_b$ and 
$A_b$, the measurements of these quantities can in principle be used
to constrain the parameter space of the models.  
The radiative corrections to $R_b$ from extended Higgs sectors
are typically of the same order of magnitude as the experimental error in the
$R_b$ measurement.  Thus $R_b$ can be used to constrain the models.
However, the radiative corrections to $A_b$ from
extended Higgs sectors are much smaller than the experimental error
in the $A_b$ measurement.  They are also much smaller than the deviation
of the $A_b$ measurement from the SM prediction.
We conclude that if $A_b \neq A_b^{\rm SM}$, the deviation does not arise
from the contributions of an extended Higgs sector.

In this paper we obtained general formulae for the corrections to the 
$Zb\bar{b}$ vertex, and then used the general formulae
to calculate the contributions to $R_b$ and $A_b$ in specific models.  
Here we summarize our conclusions for the various models.

The contributions from neutral Higgs boson exchange are only significant in a 
Type II model with enhanced $\lambda_b$.  
The regions of parameter space in which 
the contribution to $R_b$ from neutral Higgs boson exchange can be positive
is nearly ruled out by direct Higgs boson searches.
Otherwise, the contribution to $R_b$ is negative, giving a worse agreement
with experiment than the SM.  
A pair of neutral Higgs states, $H^0$ and $A^0$, with a significant $ZH^0A^0$ 
coupling and a large mass splitting, gives a large negative contribution
to $R_b$.  The $R_b$ measurement can then be used to exclude these
regions of parameter space.

The contributions to $R_b$ from charged Higgs boson exchange are negative
in models which contain only doublets and singlets, and in any model 
whose Higgs sector preserves ${\rm SU}(2)_c$ symmetry.
If the contributions from neutral Higgs boson exchange in these models 
are not significant ({\it e.g.}, if $\lambda_b$ is small), then $R_b$ 
sets a lower bound on the masses of the charged Higgs states.  
The lower bound depends on $\lambda_t$ and the charged Higgs mixing 
angles.

The contribution to $R_b$ from charged Higgs boson exchange can only be
positive if the model contains one of two features.  It must either
contain off--diagonal $ZH^+_iH^-_j$ couplings in which both of the
charged Higgs bosons couple to quarks and have different masses, or it must
contain diagonal $ZH^+_iH^-_i$ couplings which differ from the 
couplings in doublet models, or both.  This can only happen in 
models which contain Higgs multiplets larger than doublets and are
not constrained by ${\rm SU}(2)_c$ symmetry.  In such a model, the vevs 
of the multiplets larger than doublets must be very
small in order for the model to be consistent with the measured 
value of the $\rho$ parameter.
With this constraint, the contribution to $R_b$ can only be positive
when the model contains more than one doublet and there is significant
mixing between the doublets and the larger multiplets.

The precision of the $R_b$ and $A_b$ measurements is not likely to
improve significantly in the near future.  Most of the LEP and SLC $Z$
pole data has been analyzed, and no further running of these machines
at $\sqrt{s}=M_Z$ is anticipated.  Thus, future constraints on
extended Higgs sectors must come from other sources.  New virtual
constraints on extended Higgs sectors will come from measurements of
$b$ quark decays at the high-luminosity $B$-factories which will
soon be sensitive to a variety of rare $B$ decay modes. 
For example, the processes 
$b \to s \ell^+ \ell^-$ and $b\to s\gamma$ are sensitive
to the virtual charged-Higgs exchange (the latter process has already
been used to constrain the extended Higgs parameter space).
In addition, the process $b \to s
\tau^+ \tau^-$ receives a contribution from a neutral Higgs boson
coupled to the $\tau^+ \tau^-$ pair, whereas $b \to c \tau \nu$
receives a contribution from tree--level charged Higgs boson exchange
\cite{PKrawczyk88,Kalinowski90,Grossman94}.  High statistics samples
of these decay modes will yield interesting new constraints on the
structure of the extended Higgs sector.

Ultimately, one will need to directly probe the extended Higgs sector
by explicitly producing the scalar states (beyond $h^0$ which may
resemble the SM Higgs boson) at future colliders.  If some signal is
seen, it will be a demanding task to interpret
the signal and deduce the structure of the underlying scalar sector.
The constraints on the Higgs sector parameter space from $R_b$ and
other rare $B$ decay modes can play an important role in helping to
unravel the physics of the Higgs sector and probe the origin of
electroweak symmetry breaking.

\appendix
\section{Contribution to the \boldmath$\rho$ parameter in the 2HDM}
\label{app:2HDM}

In this appendix we give the one--loop contribution of the Higgs bosons 
in the 2HDM to the $\rho$
parameter, from ref.~\cite{Haber93rho}:\footnote{A typographical error
in the formula for $\Delta \rho$ in ref.~\cite{Haber93rho} is corrected
in eq.~(\ref{rho2HDM}).} 
\def\mw{M_W}
\def\mz{M_Z}
\def\hpm{H^{\pm}}
\def\mhpm{M_{\hpm}}
\def\mhp{M_{\hp}}
\def\hp{H^+}
\def\hm{H^-}
\def\hl{h^0}
\def\hh{H^0}
\def\ha{A^0}
\def\mhl{M_{\hl}}
\def\mhh{M_{\hh}}
\def\mha{M_{\ha}}
\begin{eqnarray} \label{rho2HDM}
\Delta \rho &=& \frac{\alpha}{16\pi\mw^2 s_W^2} \Biggl\{
F(\mhpm^2,\mha^2)+\sin^2(\beta-\alpha)\left[F(\mhpm^2,\mhh^2)-
F(\mha^2,\mhh^2)\right]  \nonumber \\ 
& &
+\cos^2(\beta-\alpha)\biggl[F(\mhpm^2,\mhl^2)
-F(\mha^2,\mhl^2)+F(\mw^2,\mhh^2)  \nonumber \\[5pt]
& &
\qquad\qquad -F(\mw^2,\mhl^2)
-F(\mz^2,\mhh^2)+F(\mz^2,\mhl^2) \nonumber  \\[5pt]
& & \qquad\qquad
+4\mz^2\left[B_0(0;\mz^2,\mhh^2)-B_0(0;\mz^2,\mhl^2)\right]  \nonumber \\
& & \qquad\qquad
-4\mw^2\left[B_0(0;\mw^2,\mhh^2)-B_0(0;\mw^2,\mhl^2)\right]\biggr]\Biggr\}
\end{eqnarray}
where $s_W \equiv \sin\theta_W$, and
\begin{equation}
B_0(0;m_1^2,m_2^2)={A_0(m_1^2)-A_0(m_2^2)\over m_1^2-m_2^2}
\end{equation}
\begin{equation}
A_0(m^2)\equiv m^2[\Delta+1-\log (m^2/\mu^2)]
\end{equation}
\begin{equation}
F(m_1^2,m_2^2)\equiv\nicefrac{1}{2}(m_1^2+m_2^2)- 
\frac{m_1^2 m_2^2}{m_1^2-m_2^2}\,\log\left(\frac{m_1^2}{m_2^2}\right).
\end{equation}
We have defined
$\Delta\rho$ relative to the SM where the
SM Higgs mass is taken equal to $M_{h^0}$.  With this definition,
$\Delta\rho$ is a finite quantity and is independent of the
scale, $\mu$, and the divergence, 
$\Delta\equiv 1/\epsilon-\gamma+\log(4\pi)$, of dimensional regularization.

\section{Constraints from direct Higgs searches}
\label{app:dirsearches}

In this appendix, we briefly summarize the constraints on extended
Higgs sectors resulting from the direct Higgs searches at LEP.

\subsection{Charged Higgs searches}

At LEP, charged Higgs bosons are produced via
$e^+e^-\to\gamma^*$, $Z^*\to H^+ H^-$.  The LEP analysis then assumes that
${\rm BR}(H^{+} \to c\bar{s}) + 
{\rm BR}(H^{+} \to \tau^+ \nu_{\tau})\simeq 1$.
The resulting limit obtained in ref.~\cite{lephiggs} is
$M_{H^+}>77.3$~GeV.  This mass limit would be relaxed if other
charged Higgs decay modes are significant.  In extended Higgs
models with two or more singly--charged Higgs bosons, 
we shall apply the LEP bound only to the lightest charged Higgs state.

The LEP bound also depends on the production cross section of the 
charged Higgs boson pair.  In the analysis of ref.~\cite{lephiggs}
it is assumed that the $ZH^+H^-$ coupling is that of the 2HDM:
\begin{equation} \label{zhphm}
g_{ZH^+H^-} = -\frac{e}{s_Wc_W} \left( \half - s^2_W \right).
\end{equation}
This coupling, and hence the resulting $H^+ H^-$ production cross
section, is the same as the one that arises
in models containing multiple doublets and singlets, and in the
G--M models for $H_3^{\pm}$.  The LEP charged Higgs mass bound 
is used for these models in figs.~\ref{fig:mtwohdm}, \ref{fig:D2T_0+} 
and \ref{fig:tripcustRb}.  However, eq.~(\ref{zhphm}) 
is {\it not} the same as the $ZH^+ H^-$ coupling that occurs
in models containing doublets and triplets without 
${\rm SU}(2)_c$ 
symmetry.  In models with one or two doublets and one real,
$Y=0$ triplet, the $ZH^+ H^-$ coupling is larger than in the 2HDM, and hence
the production cross section is larger.  Therefore in these models,
the charged Higgs mass bound from ref.~\cite{lephiggs} is 
a conservative bound.  This bound is used
in fig.~\ref{fig:D2T_0+} for the model with two doublets and one 
$Y=0$ triplet.  
In models with one or two doublets and
one complex, $Y=2$ triplet, the coupling is smaller than in the 2HDM.
Hence the charged Higgs boson production cross section is smaller, and the 
LEP charged Higgs mass bound is no longer valid.  This 
is the case in fig.~\ref{fig:D2T_2+}, for the model with two doublets
and one $Y=2$ triplet.

\subsection{Neutral Higgs searches in the 2HDM}

The search for neutral Higgs bosons at LEP focuses primarily on 
the SM Higgs boson and Higgs bosons of the
Minimal Supersymmetric Model (MSSM).
The SM Higgs boson is produced via
$e^+e^- \to Z^* \to Z h^0$.  In the MSSM, in addition to $Zh^0$
production, one can produce a CP-even Higgs boson in association
with a CP-odd Higgs boson via $e^+e^- \to Z^* \to h^0 A^0$.  
The MSSM Higgs sector is a 2HDM with particular relations among Higgs
sector parameters.  Thus, the MSSM Higgs mass bounds do not
immediately apply to the general 2HDM.  

From the combined LEP data taken at $\sqrt{s} = 189$ GeV, 
the lower limit on the SM Higgs mass obtained in 
ref.~\cite{lephiggs} is $M_{h^0_{\rm SM}} > 95.2$~GeV.
This bound depends primarily on the cross-section for
$e^+e^-\to Z^*\to Z h^0$ (under the assumption that the decay
branching fractions
of the $h^0$ follow roughly the pattern expected in the SM).
In the 2HDM, the $ZZh^0$ coupling is reduced from its SM value by a
factor of $\sin(\beta - \alpha)$, resulting in 
\begin{equation} \label{sig2HDM}
\sigma(e^+e^- \to Z h^0) = 
             \sigma_{\rm SM}(e^+e^- \to Z h^0) \sin^2(\beta - \alpha)\,.
\end{equation}

The LEP bound on $M_{h^0}$ in the SM is determined
by the mass value at which $\sigma_{\rm SM}(e^+e^- \to Z h^0)$
crosses the measured upper bound of $\sigma(e^+e^- \to Z h^0)$.
Using eq.~(\ref{sig2HDM}), this
can then be translated into a bound on $\sin^2(\beta - \alpha)$
as a function of $M_{h^0}$.  The resulting bound can be found in
fig.~4 of ref.~\cite{lephiggs}.  
For $\sin^2(\beta - \alpha) = 1$, the bound on $M_{h^0}$ is the 
same as in the SM, $M_{h^0} > 95.2$~GeV.  This bound is 
used in fig.~\ref{fig:Rb0_0_no}. 
For $\sin^2(\beta - \alpha) = 1/2$, the bound on $M_{h^0}$ is 
$M_{h^0}\gsim 90$ GeV.\footnote{In fig.~\ref{fig:Rb0_2_165},
a bound of $M_{h^0} > 87$~GeV is used,
corresponding to our best estimate based on LEP data prior to the
availability of fig.~4 of ref.~\cite{lephiggs}.}

In the above discussion, only the $Zh^0$ mode was considered.  For a
complete determination of the 2HDM parameter constraints, it is
necessary to include the LEP limits on $h^0 A^0$ (and $H^0 A^0$)
associated production via virtual $s$-channel $Z$-exchange.
The $Zh^0 A^0$ [$ZH^0 A^0$] coupling is proportional to
$\cos(\beta-\alpha)$ [$\sin(\beta-\alpha)$], so for fixed
$\sin(\beta-\alpha)$ one can deduce a region in the $M_{A^0}$ {\it vs.}
$M_{h^0}$ plane that is excluded by LEP data.
Unfortunately, the LEP neutral Higgs boson search data
are typically presented in the context of the
MSSM, where Higgs sector parameters are correlated.\footnote{A more
general 2HDM analysis has recently been presented by the OPAL
Collaboration \cite{opalhiggs}.  The results of this work came too late
to be included in the analysis of this paper, although we expect only
minor changes to our results.}   
For example,
at large $\tan\beta$ and values of $M_{A^0}\lsim M_Z$, one finds that
$M_{h^0}\approx M_{A^0}$ and
$\cos(\beta-\alpha)\simeq 1$.  This implies that in this region of
MSSM parameter space, the LEP search is sensitive only to $h^0 A^0$
production.  

To extract general 2HDM constraints, we proceed as follows.  
From the LEP search for $e^+ e^-\to h^0 A^0$, the LEP MSSM 
analysis \cite{lephiggs}
yields $M_{h^0}>80.7$~GeV and $M_{A^0}>80.9$~GeV.  These lower bounds
correspond roughly to pure $h^0 A^0$ production at large $\tan\beta$.
We can convert this into an upper limit for the $h^0 A^0$
cross-section for Higgs mass values at the respective lower bounds.
To get results that apply more generally to the 2HDM (where $M_{h^0}$,
$M_{A^0}$ and $\cos(\beta-\alpha)$ are not correlated), we make the
simplifying assumption that the Higgs boson detection efficiency and
background is fairly flat as a function of the Higgs masses.  We can then
vary $M_{h^0}$ and $\cos^2(\beta - \alpha)$ and find a lower bound 
on $M_{A^0}$.\footnote{For $|\cos(\beta-\alpha)| \ll 1$, the production
of $H^0 A^0$ rather than $h^0 A^0$ is relevant.}  
The resulting ``direct search'' bounds have been implemented 
in figs.~\ref{fig:Rb0_2_165} and \ref{fig:Rb0_1_165}.  Further details of
this analysis can be found in ref.~\cite{hlthesis}.

\end{document}